\documentclass[prd,nofootinbib,twocolumn,superscriptaddress,preprintnumbers,balancelastpage,longbibliography]{revtex4-1}
\usepackage{ae,aecompl}
\usepackage{graphicx}	
\usepackage{amsmath}	
\usepackage{graphicx}
\usepackage{dcolumn}
\usepackage{bm}
\usepackage{graphics}
\usepackage{afterpage}
\usepackage{float}
\usepackage{subfigure}
\usepackage{rotating}
\usepackage{multirow}
\usepackage{fancyheadings}
\usepackage{theorem}
\usepackage{moreverb}
\usepackage{euscript}
\usepackage{psfrag}
\usepackage{slashed}
\usepackage{mathtools}
\usepackage[colorlinks=true
,urlcolor=blue
,anchorcolor=blue
,citecolor=blue
,filecolor=blue
,linkcolor=blue
,menucolor=blue
,linktocpage=true
,pdfproducer=medialab
,pdfa=true
]{hyperref}
\usepackage{listings}
\usepackage{color,xcolor}

\definecolor{deepblue}{rgb}{0,0,0.9}
\definecolor{deepred}{rgb}{0.85,0,0}
\definecolor{deepgreen}{rgb}{0,0.95,0}

\lstdefinestyle{python}{
  belowcaptionskip=1\baselineskip,
  breaklines=true,
  frame=L,
  xleftmargin=\parindent,
  language=Python,
  showstringspaces=false,
  basicstyle=\small\ttfamily,
  morekeywords={models, lambda, forms,True,False,None},
  keywordstyle=\bfseries\color{deepgreen!40!black},
  commentstyle=\itshape\color{gray},
  identifierstyle=\color{black},
  stringstyle=\color{deepred},
  rulecolor=\color{gray},
}

\newcommand{\es}[2] {\begin{equation} \label{#1} \begin{split} #2 \end{split} \end{equation}}

\begin{document}

\title{\texttt{NPTFit}: A code package for Non-Poissonian Template Fitting}
\author{Siddharth Mishra-Sharma}
\email{smsharma@princeton.edu}
\affiliation{Department of Physics, Princeton University, Princeton, NJ 08544}
\author{Nicholas L. Rodd}
\email{nrodd@mit.edu}
\affiliation{Center for Theoretical Physics, Massachusetts Institute of Technology, Cambridge, MA 02139}
\author{Benjamin R. Safdi}
\email{bsafdi@mit.edu}
\affiliation{Center for Theoretical Physics, Massachusetts Institute of Technology, Cambridge, MA 02139}
\preprint{MIT-CTP/4864, PUPT 2515}
\date{\today}

\begin{abstract}
We present \texttt{NPTFit}, an open-source code package, written in \texttt{python} and \texttt{cython}, for performing non-Poissonian template fits (NPTFs).  The NPTF is a recently-developed statistical procedure for characterizing the contribution of unresolved point sources (PSs) to astrophysical data sets.  The NPTF was first applied to {\it Fermi} gamma-ray data
to give evidence that the excess of $\sim$GeV gamma-rays observed in the inner regions of the Milky Way likely arises from a population of sub-threshold point sources, and the NPTF has since found additional applications studying sub-threshold extragalactic sources at high Galactic latitudes.  The NPTF generalizes traditional astrophysical template fits to allow for the ability to search for populations of unresolved PSs that may follow a given spatial distribution. \texttt{NPTFit} builds upon the framework of the fluctuation analyses developed in X-ray astronomy, and thus likely has applications beyond those demonstrated with gamma-ray data.  The \texttt{NPTFit} package utilizes novel computational methods to perform the NPTF efficiently. The code is available at  \url{https://github.com/bsafdi/NPTFit} and up-to-date and extensive documentation may be found at \url{http://nptfit.readthedocs.io}.
\end{abstract}

\maketitle

\section{Introduction}

Astrophysical point sources (PSs), which are defined as sources with angular extent smaller than the resolution of the detector, play an important role in virtually every analysis utilizing images of the cosmos.  It is useful to distinguish between resolved and unresolved PSs; the former may be detected individually at high significance, while members of the latter population are by definition too dim to be detected individually.  However, unresolved PSs -- due to their potentially large number density -- can be a leading and sometimes pesky source of flux across wavelengths.
  Recently, a novel analysis technique called the non-Poissonian template fit (NPTF) has been developed for characterizing populations of unresolved PSs at fluxes below the detection threshold for finding individually-significant sources~\cite{Lee:2014mza,Lee:2015fea}.  The technique expands upon the traditional fluctuation analysis technique (see, for example,~\cite{Miyaji:2001dp,Malyshev:2011zi}), which analyzes the aggregate photon-count statistics of a data set to characterize the contribution from unresolved PSs, by additionally incorporating spatial information both for the distribution of unresolved PSs and for the potential sources of non-PS emission.  In this work, we present a code package called \texttt{NPTFit} for numerically implementing the NPTF in \texttt{python} and \texttt{cython}.
  
The most up-to-date version of the open-source package \texttt{NPTFit} may be found at
\begin{center}
\url{https://github.com/bsafdi/NPTFit}
\end{center}
and the latest documentation at
\begin{center}
\url{http://nptfit.readthedocs.io}.
\end{center}
In addition, the version used in this paper has been archived at
\begin{center}
\url{https://zenodo.org/record/380469#.WN_pSFPyvMV}.
\end{center}

The NPTF generalizes traditional astrophysical template fits.  Template fitting is useful for pixelated data sets consisting of some number of photon counts $n_p$ in each pixel $p$, and it typically proceeds as follows.   Given a set of model parameters $\bm{\theta}$, the mean number of predicted photon counts $\mu_p({\bm{\theta}})$ in the pixel $p$ may be computed.  More specifically, $\mu_p({\bm{\theta}}) = \sum_{\ell} T^{(S)}_{p,\ell}({\bm{\theta}})$, where $\ell$ is an index of the set of templates $T^{(S)}_{p, \ell}$, whose normalizations and spatial morphologies may depend on the parameters $\bm{\theta}$. These templates may, for example, trace the gas-distribution or other extended structures that are expected to produce photon counts.  Then, the probability to detect $n_p$ photons in the pixel $p$ is simply given by the Poisson distribution with mean $\mu_p({\bm{\theta}})$.  By taking a product of the probabilities over all pixels, it is straightforward to write down a likelihood function as a function of ${\bm{\theta}}$.  

The NPTF modifies this procedure by allowing for non-Poissonian photon-count statistics in the individual pixels.  That is, unresolved PS populations are allowed to be distributed according to spatial templates, but in the presence of unresolved PSs the photon-count statistics in individual pixels, as parameterized by ${\bm{\theta}}$, no longer follow Poisson distributions.  This is heuristically because we now have to ask two questions in each pixel: first, what is the probability, given the model parameters ${\bm{\theta}}$ that now also characterize the intrinsic source-count distribution of the PS population, that there are PSs within the pixel $p$, then second, given that PS population, what is the probability to observe $n_p$ photons?

It is important to distinguish between resolved and unresolved PSs.  Once a PS is resolved -- that is once its location and flux is known -- that PS may be accounted for by its own Poissonian template.  Unresolved PSs are different because their locations and fluxes are not known.  When we characterize unresolved PSs with the NPFT, we characterize the entire population of unresolved sources, following a given spatial distribution, based on how that population modifies the photon-count statistics.
  
The NPTF has played an important role recently in addressing various problems in gamma-ray astroparticle physics with data collected by the \emph{Fermi}-LAT gamma-ray telescope.\footnote{\url{http://fermi.gsfc.nasa.gov/}} The NPTF was developed to address the excess of gamma rays observed by \emph{Fermi} at $\sim$GeV energies originating from the inner regions of the Milky Way~\cite{Goodenough:2009gk,Hooper:2010mq,Boyarsky:2010dr,Hooper:2011ti,Abazajian:2012pn,Hooper:2013rwa,Gordon:2013vta,Abazajian:2014fta,Daylan:2014rsa,Calore:2014xka,Abazajian:2014hsa,TheFermi-LAT:2015kwa,Macias:2016nev,Clark:2016mbb}.  The GeV excess, as it is commonly referred to, has received a significant amount of attention due to the possibility that the excess emission arises from dark matter (DM) annihilation.  However, it is well known that unresolved PSs may complicate searches for annihilating DM in the Inner Galaxy region due to, for example, the expected population of dim pulsars~\cite{Abazajian:2014fta,Abazajian:2010zy,Hooper:2013nhl,Calore:2014oga,Cholis:2014lta,Petrovic:2014xra,Yuan:2014yda,OLeary:2015gfa,Brandt:2015ula}.  In~\cite{Lee:2015fea} (see also~\cite{Linden:2016rcf}) it was shown, using the NPTF, that indeed the photon-count statistics of the data prefer a PS over a smooth DM interpretation of the GeV excess.  The same conclusion was also reached by~\cite{Bartels:2015aea} using an unrelated method that analyzes the statistics of peaks in the wavelet transformation of the \emph{Fermi} data.  

In the case of the GeV excess, there are multiple PS populations that may contribute to the observed gamma-ray flux and complicate the search for DM annihilation.  These include isotropically distributed PSs of extragalactic origin, PSs distributed along the disk of the Milky Way such as supernova remnants and pulsars, and a potential spherical population of PSs such as millisecond pulsars.  Additionally, there are various identified PSs that contribute significantly to the flux as well as a variety of smooth emission mechanisms such as gas-correlated emission from pion decay and bremsstrahlung.  The power of the NPTF is that these different source classes may be given separate degrees of freedom and constrained by incorporating the spatial morphology of their various contributions along with the difference in photon-count statistics between smooth emission and emission from unresolved PSs.  Although the origin of the GeV excess is still not completely settled, as even if the excess arises from PSs as the NPTF suggests the source class of the PSs remains a mystery at present, the NPTF has emerged as a powerful tool for analyzing populations of dim PSs in complicated data sets with characteristic spatial morphology.

The NPTF and related techniques utilizing photon-count statistics have also been used recently to study the contribution of various source classes to the extragalactic gamma-ray background (EGB)~\cite{Malyshev:2011zi,Zechlin:2015wdz,TheFermi-LAT:2015ykq,Zechlin:2016pme,Lisanti:2016jub}.\footnote{The complementary analysis strategy of probabilistic catalogues has also been applied to this problem \cite{Daylan:2016tia}.}  In these works it was shown that unresolved blazars would predominantly show up as PS populations under the NPTF, while other source classes such as star-forming galaxies would show up predominantly as smooth emission.  For example, in~\cite{Lisanti:2016jub} it was shown using the NPTF that blazars likely account for the majority of the EGB from $\sim$2 GeV to $\sim$2 TeV.  These results set strong constraints on the flux from more diffuse sources, such as star-forming galaxies, which has significant implications for, among other problems, the interpretation of the high-energy astrophysical neutrinos observed by IceCube~\cite{Aartsen:2013bka,Aartsen:2013jdh,Aartsen:2015knd,Aartsen:2015rwa} (see, for example,~\cite{Bechtol:2015uqb,Murase:2016gly}).  This is because certain sources that contribute gamma-ray flux at {\it Fermi} energies, such as star forming galaxies and various types of active galactic nuclei, may also contribute neutrino flux observable by IceCube.   

The NPTF originates from the older fluctuation analysis technique, which is sometimes referred to as the $P(D)$ analysis.  This technique has been used extensively to study the flux of unresolved X-ray sources~\cite{hasinger1993,1993MNRAS.262..619G,Gendreau:1997di,Perri:2000fv,Miyaji:2001dp}.  In these early works, the photon-count probability distribution function (PDF) was computed numerically for different PS source-count distributions using Monte Carlo (MC) techniques.  The fluctuation analysis was first applied to gamma-ray data in~\cite{Malyshev:2011zi},\footnote{The fluctuation analysis has more recently been applied to both gamma-ray \cite{Feyereisen:2015cea} and neutrino \cite{Feyereisen:2016fzb} datasets.} and in that work the authors developed a semi-analytic technique utilizing probability generating functions for calculating the photon-count PDF.  The code package $\texttt{NPTFit}$ presented in this work uses this formalism for efficiently calculating the photon-count PDF.  The specific form of the likelihood function for the NPTF, while reviewed in this work, was first presented in~\cite{Lee:2015fea}.  The works~\cite{Lee:2015fea,Linden:2016rcf,Lisanti:2016jub} utilized an early version of $\texttt{NPTFit}$ to perform their numerical analyses. 

The $\texttt{NPTFit}$ code package has a \texttt{python} interface, though the likelihood evaluation is efficiently implemented in \texttt{cython}~\cite{behnel2010cython}.  The user-friendly interface allows for an arbitrary number of PS and smooth templates.  The PS templates are characterized by pixel-dependent source-count distributions $dN_p/dF = T_p^{({\rm PS})} dN/dF$, where $T_p^{({\rm PS})}$ is the spatial template tracking the distribution of point sources on the sky and $dN/dF$ is the pixel-independent source-count distribution.  The distribution $dN_p/dF$ quantifies the number of sources $dN_p$ that contributes flux between $F$ and $F + dF$ in the pixel $p$.  The $dN/dF$ are parameterized as multiply broken power-laws, with an arbitrary number of breaks.  The code is able to account for both an arbitrary exposure map (accounting for the pointing strategy of an instrument) as well as an arbitrary point spread function (PSF, accounting for the instrument's finite angular resolution) in translating between flux $F$ and counts $S$. 

 \texttt{NPTFit} has a built-in interface with \texttt{MultiNest}~\cite{Feroz:2008xx,Buchner:2014nha}, which efficiently implements nested sampling of the posterior distribution and Bayesian evidence for the user-specified model, given the specified data and instrument response function, in the Bayesian framework~\cite{Feroz:2013hea,Feroz:2007kg,skilling2006}.  The interface handles the Message Passing Interface (MPI), so that inference may be performed efficiently using parallel computing.  A basic analysis package is provided in order to facilitate easy extraction of the most relevant data from the posterior distribution and quick plotting of the \texttt{MultiNest} output.  The preferred format of the data for \texttt{NPTFit} is \texttt{HEALPix}~\cite{Gorski:2004by} (a nested equal-area pixilation scheme of the sky), although the the code is also able to handle non-\texttt{HEALPix} data arrays. Note that the code package may also be used to simply extract the NPTF likelihood function so that \texttt{NPTFit} may be interfaced with any numerical package for Bayesian or frequentist inference.

A large set of example \texttt{Jupyter}~\cite{PER-GRA:2007} notebooks and \texttt{python} files are provided to illustrate the code.  The examples utilize 413 weeks of processed \emph{Fermi} Pass 8 data in the UltracleanVeto event class collected between August 4, 2008 and July 7, 2016 in the energy range from 2 to 20 GeV. We restrict this dataset to the top quartile as graded by PSF reconstruction and further apply the standard quality cuts \texttt{DATA\_QUAL==1 \&\& LAT\_CONFIG==1}, as well as restricting the zenith angle to be less than $90^\circ$. This data is made available in the code release.  Moreover, the example notebooks illustrate many of the main results in~\cite{Lee:2015fea,Linden:2016rcf,Lisanti:2016jub}.

In addition to the above, the base \texttt{NPTFit} code makes use of the \texttt{python} packages \texttt{corner}~\cite{dan_foreman_mackey_2016_53155}, \texttt{matplotlib}~\cite{Hunter:2007}, \texttt{mpmath}~\cite{mpmath}, \texttt{GSL}~\cite{galassi2015gnu} and  \texttt{numpy}~\cite{oliphant2006guide}.

The rest of this paper is organized as follows.  Section~\ref{NPTF} outlines in more detail the framework of the NPTF.  Section~\ref{NPTFit-orientation} highlights the key classes and features in the \texttt{NPTFit} code package and usage instructions.  In Sec.~\ref{NPTFit-example} we present an example of how to perform an NPTF scan using \texttt{NPTFit}, looking at the Galactic Center with \emph{Fermi} data to reproduce aspects of the main results of~\cite{Lee:2015fea}. We conclude in Sec.~\ref{Conclusion}. Appendices~\ref{details},~\ref{algorithms}, and~\ref{xmcalc} describe further details behind the mathematical framework of the NPTF.      

\section{The Non-Poissonian Template Fit}
\label{NPTF}

In this section we review the NPTF, which was first presented in~\cite{Lee:2015fea} and described in more detail in~\cite{Linden:2016rcf,Lisanti:2016jub} (see also~\cite{Malyshev:2011zi,Lee:2014mza,Zechlin:2015wdz,Zechlin:2016pme}).  The NPTF is used to fit a model $\mathcal{M}$ with parameters $\bm{\theta}$ to a data set $d$ consisting of counts $n_p$ in each pixel $p$.  The likelihood function for the NPTF is then simply
\es{eq:likelihood}{
p(d |{\bm \theta}, \mathcal{M}) = \prod_p p_{n_p}^{(p)}({\bm \theta}) \,,
}
where $p_{n_p}^{(p)}( {\bm \theta})$ gives the probability of drawing $n_p$ counts in the given pixel $p$, as a function of the parameters $\bm{\theta}$.  The main computational challange, of course, is in computing these probabilities.

It is useful to divide the model parameters into two different categories: the first category describes smooth templates, while the second category describes PS templates.  We describe each category in turn, starting with the smooth templates.  

For most applications, the data has the interpretation of being a two-dimensional pixelated map consisting of an integer number of counts in each pixel.  The smooth templates may be used to predict the mean number of counts $\mu_p({\bm \theta})$ in each pixel $p$:   
\es{mean_pixel}{
\mu_p( {\bm \theta}) = \sum_\ell \mu_{p, \ell} ({\bm \theta}) \,.
}
Above, $\ell$ is an index over templates and $\mu_{p, \ell} ({\bm \theta})$ denotes the mean contribution of the $\ell^\text{th}$ template to pixel $p$ for parameters ${\bm \theta}$.  In principle, ${\bm \theta}$ may describe both the spatial morphology as well as the normalization of the templates.  However, in the current implementation of the code, the Poissonian model parameters simply characterize the overall normalization of the templates: $\mu_{p, \ell} ({\bm \theta}) = A_{\ell}( {\bm \theta}) T^{(S)}_{p,\ell}$.  Here, $A_{\ell}$ is the normalization parameter and $T^{(S)}_{p,\ell}$ is the $\ell^\text{th}$ template, which takes values over all pixels $p$ and is independent of the model parameters.  The superscript $(S)$ implies that the template is a counts templates, which is to be contrasted with a flux template, for which we use the symbol $(F)$. The two are related by the exposure map of the instrument $E_{p}$: $T^{(S)}_p = E_p T^{(F)}_p$. In the case where we only have smooth, Poissonian templates, the probabilities are then given by the Poisson distribution:
\es{poisson}{
p_{n_p}^{(p)}({\bm \theta}) = {\mu_p^{n_p}( {\bm \theta}) \over n_p !} e^{- \mu_p( {\bm \theta}) } \,.
}

In the presence of unresolved PS templates, the probabilities $p_{n_p}^{(p)}({\bm \theta})$ are no longer Poissonian functions of the model parameters ${\bm \theta}$.  Each PS template is characterized by a pixel-dependent source-count distribution $dN_p/dF$, which describes the differential number of sources per pixel per unit flux interval.  In this work, we model the source-count distribution by a multiply broken power-law:  
\begin{widetext}
\es{mbpl}{
\frac{dN_p}{dF} (F; {\bm \theta}) = A ( {\bm \theta}) T^{({\rm PS})}_p \left\{ \begin{array}{lc} \left( \frac{F}{F_{b,1}} \right)^{-n_1}, & F \geq F_{b,1} \\ \left(\frac{F}{F_{b,1}}\right)^{-n_2}, & F_{b,1} > F \geq F_{b,2} \\ \left( \frac{F_{b,2}}{F_{b,1}} \right)^{-n_2} \left(\frac{F}{F_{b,2}}\right)^{-n_3}, & F_{b,2} > F \geq F_{b,3} \\ \left( \frac{F_{b,2}}{F_{b,1}} \right)^{-n_2} \left( \frac{F_{b,3}}{F_{b,2}} \right)^{-n_3} \left(\frac{F}{F_{b,3}}\right)^{-n_4}, & F_{b,3} > F \geq F_{b,4} \\ \\
\ldots & \ldots \\ \\
\left[ \prod_{i=1}^{k-1} \left( \frac{F_{b,i+1}}{F_{b,i}} \right)^{-n_{i+1}} \right] \left( \frac{F}{F_{b,k}} \right)^{-n_{k+1}}, & F_{b,k} > F \end{array} \right. .
}
\end{widetext}
Above, we have parameterized the source-count distribution with an arbitrary number of breaks $k$, denoted by $F_{b,i}$ with \mbox{$i \in [1,2, \ldots, k]$}, and $k+1$ indices $n_i$ with \mbox{$i \in [1,2, \ldots , k+1]$}.  The spatial dependence of the source-count distribution is accounted for by the overall factor $A ( {\bm \theta}) T_p^{({\rm PS})}$, where $A ( {\bm \theta})$ is the pixel-independent normalization, which is a function of the model parameters, and $T_p^{({\rm PS})}$ is a template describing the spatial distribution of the PSs.  More precisely, the number of sources $N^\text{PS}_p = \int dF dN_p / dF$ (and the total PS flux $F^\text{PS}_p = \int dF F dN_p / dF$) in pixel $p$, for a fixed set of model parameters ${\bm \theta}$, follows the template $T_p^{({\rm PS})}$.  On the other hand, the locations of the flux breaks and the indices are taken to be fixed between pixels.\footnote{In principle, the breaks and indices could also vary between pixels.  However, in the current version of \texttt{NPTFit}, only the number of sources (and, accordingly, the total flux) is allowed to vary between pixels.} 

To summarize, a PS template described by a broken power-law with $k$ breaks has $2 (k+1)$ model parameters describing the locations of the breaks, the power-law indices, and the overall normalization.  For example, if we take a single break then the PS model parameters may be denoted as $\{ A, F_{b,1}, n_1, n_2 \}$.  Additionally, a spatial template $T^{({\rm PS})}$ must be specified, which describes the distribution of the number of sources (and total flux) with pixel $p$.     

Notice that when we discussed the Poissonian templates we used the counts templates $T^{(S)}$ and talked directly in terms of counts $S$, while so far in our discussion of the unresolved PS templates we have used the point source distribution template $T^{({\rm PS})}$ and written the source-count distribution $dN/dF$ in terms of flux $F$. Of course as the total flux from a distribution of point sources is also proportional to the template $T^{({\rm PS})}$, it can be thought of as a flux template, however conceptually it is being used to track the distribution of the sources rather than the flux they produce. For this reason we have chosen to distinguish the two. Moreover, in the presence of a non-trivial PSF, $T^{(S)}$ should also be smoothed by the PSF to account for the instrument response function.  That is, $T^{(S)}$ is a template for the observed counts taking into account the details of the instrument, while $T^{({\rm PS})}$ ($T^{(F)}$) is a map of the physical point sources (flux), which is independent of the instrument.  In photon-counting applications, the exposure map $E_p$ often has units of $\text{cm}^2 \text{s}$ and flux has units of $\text{counts}\, \text{cm}^{-2}\text{s}^{-1}$.

For the unresolved PS templates, we also need to convert the source-count distribution from flux to counts.  This is done by a simple change of variables:
\es{dNdS}{
{dN_p \over dS} (S; {\bm \theta}) = \frac{1}{E_p} {dN_p \over dF} (F = S / E_p; {\bm \theta}) \,,
}
which implies that for a non-Poissonian template the spatial dependence of $dN_p/dS$ is given by $T^{({\rm PS})}_p/E_p$. This inverse exposure scaling may seem surprising, but it is straightforward to confirm that the mean number of counts in a given pixel, $ \int dS S dN_p / dS$, is given by $E_p T^{({\rm PS})}_p$, as expected, up to pixel independent factors.

As an important aside, the template $T^{(S)}$ used by the Poissonian models needs to be smoothed by the PSF.  Incorporating the PSF into the unresolved PS models, on the other hand, is more complicated and is not accomplished simply by smoothing the spatial template.  Indeed, $T^{({\rm PS})}_p$ should remain un-smoothed by the PSF when used for non-Poissonian scans.

In the remainder of this section we briefly overview the mathematic framework behind the computation of the $p_{n_p}^{(p)}({\bm \theta})$ with \texttt{NPTFit}; however, details of the algorithms used to calculate these probabilities in practice, along with more in-depth explanations, are given in Apps.~\ref{details},~\ref{algorithms}, and~\ref{xmcalc}.  We use the probability generating function formalism, following~\cite{Malyshev:2011zi}, to calculate the probabilities.   
For a discrete probability distribution $p_k$, with $k=0,1,2,\ldots$, the generating function is defined as:
\begin{equation}
P(t) \equiv \sum_{k=0}^{\infty} p_k t^k \,,
\label{prob-gen}
\end{equation}
from which we can recover the probabilities:
\es{deriv}{
p_k = \frac{1}{k!} \left. \frac{d^k P(t)}{dt^k} \right|_{t=0} \,.
}
The key feature of generating functions exploited here is that the generating function of a sum of two independent random variables is simply the product of the individual generating functions. 

The probability generating function for the smooth templates, as a function of ${\bm \theta}$, is simply given by
\es{P-PGF}{
P_{\rm P}(t; {\bm \theta}) = \prod_p \text{exp}\left[ \mu_p( {\bm \theta}) (t - 1) \right]  \,.
}
The probability generating function for an unresolved PS template, on the other hand, takes a more complicated form:
\es{NP-PGF}{
P_{\rm NP}(t; {\bm \theta}) = \prod_p \exp \left[ \sum_{m=1}^{\infty} x_{p,m}( {\bm \theta}) ( t^m - 1) \right] \,,
}
where
\es{xm-def}{
x_{p,m}( {\bm \theta}) =\int_0^{\infty} dS \frac{dN_p}{dS}(S;{\bm \theta}) \int_0^1 df \rho(f) \frac{(fS)^m}{m!} e^{-fS} \,.
}
Above, $\rho(f)$ is a function that takes into account the PSF, which we describe in more detail in App.~\ref{details}.  In the presence of a non-trivial PSF, the flux from a single source is smeared among pixels.  The distribution of flux fractions among pixels is described by the function $\rho(f)$, where $f$ is the flux fraction.  By definition $\rho(f) df$ equals the number of pixels which, on average, contain between $f$ and $f+ df$ of the flux from a PS; the distribution is normalized such that $ \int_0^1 df f \rho(f) = 1$.  If the PSF is a $\delta$-function, then $\rho(f) = \delta(f-1)$.

Putting aside the PSF correction for the moment, the $x_{p,m}$ have the interpretation of being the average number of $m$-count PSs within the pixel $p$, given the distribution $ dN_p(S;{\bm \theta})/dS $.  The generating function for $x_m$ $m$-count sources is simply $e^{x_m(t^m - 1)}$ (see~\cite{Malyshev:2011zi} or App.~\ref{details}), which then leads directly to~\eqref{NP-PGF}.  The PSF correction, through the distribution $\rho(f)$, incorporates the fact that PSs only contribute some fraction of their flux within a given pixel.

\section{\texttt{NPTFit}: orientation}
\label{NPTFit-orientation}

\texttt{NPTFit} implements the NPTF, as described above, in \texttt{python}.  In this section we give a brief orientation to the code package and its main classes.  A more thorough description of the code and its uses is available in the \href{http://nptfit.readthedocs.io}{online documentation}.

\subsection*{ \lstinline{class NPTFit.nptfit.NPTF} }

This is the main class used to set up and perform non-Poissonian and Poissonian template scans.  It is initialized by   
\begin{lstlisting}
nptf = NPTF(tag='Untagged',work_dir=None)
\end{lstlisting}
with keywords 
\begin{center}
\begin{tabular}{ |c|c|c|c |}
\hline
Argument &  Default & Purpose & \lstinline!type! \\ \hline \hline
\lstinline!tag! & \lstinline!'Untagged'! & Label of scan & \lstinline!str! \\ \hline 
\lstinline!work_dir! & \lstinline!None! & Output directory & \lstinline!str!  \\ \hline 
\end{tabular} \,.
\end{center}
If no \lstinline{work_dir} is specified, the code will default to the current directory.  This is the directory where all output is stored.  Specifying a \lstinline{tag} will create an additional folder, with that name, within the \lstinline{work_dir} for the output.

The data, exposure map, and templates are loaded into the \lstinline{nptfit.NPTF} instance after initialization (see the example in Sec.~\ref{NPTFit-example}).  The data and exposure map are loaded by 
\begin{lstlisting}
nptf.load_data(data, exposure)
\end{lstlisting}
Here, \lstinline{data} and \lstinline{exposure} are 1-D \texttt{numpy} arrays.  The recommended format for these arrays is the \texttt{HEALPix} format, so that all pixels are equal area, although the code is able to handle arbitrary data and exposure arrays so long as they are of the same length.  The templates are added by 
\begin{lstlisting}
nptf.add_template(template, key, 
    units='counts')
\end{lstlisting}  
Here, \lstinline{template} is a 1-D  \texttt{numpy} array of the same length as the data and exposure map, \lstinline{key} is a string that will be used to refer to the template later on, and \lstinline{units} specifies whether the template is a counts template (keyword \lstinline{'counts'}) or a flux template (keyword \lstinline{'flux'}) in units  $\text{counts}\, \text{cm}^{-2}\text{s}^{-1}$.  The default, if unspecified, is \lstinline{units = 'counts'}.  The template should be pre-smoothed by the PSF if it is going to be used for a Poissonian model.  If the template is going to be used for a non-Poissonian model, either choice for \lstinline{units} is acceptable, though in the case of \lstinline{'counts'} the template should simply be the product of the exposure map times the flux template and not smoothed by the PSF. 

The user also has the option of loading in a mask that reduces the region of interest (ROI) to a subset of the pixels in the data, exposure, and template arrays.  This is done through the command
\begin{lstlisting}
nptf.load_mask(mask)
\end{lstlisting}
where \lstinline{mask} is a boolean \texttt{numpy} array of the same length as the data and exposure arrays.  Pixels in \lstinline{mask} should be either \texttt{True} or \texttt{False}; by convention, pixels that are \texttt{True} will be masked, while those that are \texttt{False} will not be masked.  Note if performing an analysis with non-Poissonian templates, regions where the exposure map is identically zero should be explicitly masked. 

Afterwards, Poissonian and non-Poissonian models may be added to the instance using the available templates. An arbitrary number of Poissonian and non-Poissonian models may be added to the scan. Moreover, each non-Poissonian model may be specified in terms of a multiply broken power law with a user-specified number of breaks, as in~\eqref{mbpl}.  

Poissonian models are added sequentially using the syntax
\begin{lstlisting}
nptf.add_poiss_model(template_name, model_tag, prior_range=[], log_prior=False, fixed=False, fixed_norm=1.0)
\end{lstlisting}
where the keywords are
\vspace{+0.02in}
\begin{center}
\begin{tabular}{ |c|c|c|c |}
\hline
Argument &  Default & Purpose & \lstinline!type! \\ \hline \hline
\lstinline!template_name! & - & \lstinline!key! of template & \lstinline!str! \\ \hline 
\lstinline!model_tag! & - & \LaTeX-ready label & \lstinline!str!  \\ \hline 
\lstinline!prior_range! & \lstinline![]! & Prior [min, max ] & [\lstinline!float!, \lstinline!float!]  \\ \hline 
\lstinline!log_prior! & \lstinline!False! & Log/linear-flat prior & \lstinline!bool!  \\ \hline 
\lstinline!fixed! & \lstinline!False! & Is template fixed & \lstinline!bool!  \\ \hline 
\lstinline!fixed_norm! & \lstinline!1.0! & Norm if  \lstinline!fixed! & \lstinline!float!  \\ \hline
\end{tabular} 
\end{center}
Any of the model parameters may be fixed to a user specified value instead of floated in the scan.  For those parameters that are floated in the scan, a prior range needs to be specified along with whether or not the prior is flat or log-flat.
Note that if \lstinline{log_prior = True}, then the prior range is set with respect to $\log_{10}$ of the linear prior range.\footnote{More complicated priors will be incorporated in future releases of \texttt{NPTFit}.}  For example, if we want to scan the normalization of a template over the range from $[0.1,10]$ with a log-flat prior, then we would set \lstinline{log_prior = True} and \lstinline{prior_range = [-1,1]}.  In this case, it might make sense to label the model with {\footnotesize\ttfamily \textcolor{black}{model\_tag =}} {\footnotesize\ttfamily \textcolor{deepred}{'\$\textbackslash log\_\{10\}A\$'}}  to emphasize that the actual model parameter is the log of the normalization; 
this label will appear in various plots made using the provided analysis class for visualizing the posterior.

The non-Poissonian models are added with a similar syntax:
\begin{lstlisting}
nptf.add_non_poiss_model(template_name, model_tag, prior_range=[], log_prior=False, dnds_model='specify_breaks', fixed_params=None, units='counts')
\end{lstlisting} 
The \lstinline{template_name} keyword is the same as for the Poissonian models.  The rest of the keywords are
\begin{widetext}
\begin{center}
\begin{tabular}{ |c|c|c|c |}
\hline
Argument &  Default & Purpose &\lstinline!type! \\ \hline \hline
\lstinline!model_tag! & - & \LaTeX-ready label & \lstinline![str, str, ...]! \\ \hline 
\lstinline!prior_range! & \lstinline![]! & Prior [[min, max], ...] & \lstinline![[float, float], ...]!  \\ \hline 
\lstinline!log_prior! & \lstinline![False]! & Log/linear-flat prior & \lstinline![bool,bool, ...]!  \\ \hline 
\lstinline!dnds_model! & \lstinline!'specify_breaks'! & How to specify multiple breaks & \lstinline!str!  \\ \hline 
\lstinline!fixed_params! & \lstinline!None! & Fix certain parameters & \lstinline![[int,float], ...]!  \\ \hline 
\lstinline!units! & \lstinline!'counts'! & \lstinline!'flux'! or \lstinline!'counts'! units for breaks  & \lstinline!str!  \\ \hline
\end{tabular} 
\end{center}
\end{widetext}
The syntax for adding non-Poissonian models is that the model parameters are specified by $[A, n_1, n_2, \ldots, n_{k+1}, S_{b,1}, S_{b,2}, \ldots, S_{b,k}]$ for a broken power-law with $k$ breaks.  As such, the \lstinline{model_tag}, \lstinline{prior_range}, and \lstinline{log_prior} are now arrays where each entry refers to the respective model parameter.  The code automatically determines the number of breaks by the length of the \lstinline{model_tag} array.  The arrays \lstinline{prior_range} and \lstinline{log_prior} should only include entries for model parameters that will be floated in the scan.  Any model parameter may be fixed using the \lstinline{fixed_params} array, with the syntax such that \lstinline{fixed_params = [[i,c_i],[j,c_j]]} would fix the $i^\text{th}$ model parameter to $c_i$ and the $j^\text{th}$ to $c_j$, where the parameter indexing starts from 0.     

The \lstinline{units} keyword determines whether the priors for the breaks in the source-count distribution (and also the fixed parameters, if any are given) will be specified in terms of \lstinline{'flux'} or \lstinline{'counts'}.  The relation between flux and counts varies between pixels if the exposure map is non-trivial.  For this reason, it is more appropriate to think of the breaks in the source-count distribution in terms of flux.  The keyword \lstinline{'counts'} still specifies the breaks in the source-count distribution in terms of flux, with the relation between counts and flux given through the mean of the exposure map $\text{mean}(E)$: $F_{b,i} = S_{b,i} / \text{mean}(E)$.  

The \lstinline{dnds_model} keyword has the options \mbox{\lstinline{'specify_breaks'}} and  \lstinline{'specify_relative_breaks'}.  If \lstinline{'specify_breaks'} is chosen, which is the default, then the breaks are the model parameters.  If instead \mbox{\lstinline{'specify_relative_breaks'}} is chosen, the full set of model parameters is given by $[A, n_1, n_2, \ldots, n_{k+1}, S_{b,1}, \lambda_{2}, \ldots, \lambda_{k}]$.  Here, $S_{b,1}$ is the highest break and the lower breaks are determined by $S_{b,i} = \lambda_i S_{b,i-1}$.  Note that the prior ranges for the $\lambda$'s should be between $0$ and $1$ (for linear flat), since $S_{b,i} < S_{b,i-1}$.  

After setting up a scan, the configuration is finished by executing the command 
\begin{lstlisting}
nptf.configure_for_scan(f_ary=[1.0], df_rho_div_f_ary=[1.0], nexp=1)
\end{lstlisting}
For a purely Poissonian scan, none of the keywords above need to be specified.  For non-Poissonian scans, \lstinline{f_ary} and \lstinline{df_rho_div_f_ary} incorporate the PSF correction.  In particular, \lstinline{f_ary} is a discretized list of $f$ values between $0$ and $1$, while  \lstinline{df_rho_div_f_ary} is a discretized list of $df \rho(f) / f$ at those $f$ values.  A class is provided for computing these lists; it is described later in this section.  If no keywords are given for these two arrays they default to the case of a $\delta$-function PSF.  

The keyword \lstinline{nexp}, which defaults to $1$, is related to the exposure correction in the calculation of the source-count distribution $dN_p/dS$ from $dN_p/dF$.  In many applications, it is computationally too expensive to perform the mapping in~\eqref{dNdS} in each pixel.  The overall pixel-dependent normalization factor $T_p^{({\rm PS})} / E_p$ factorizes from many of the internal computations, and as a result this contribution to the exposure correction is performed in every pixel.  However, it is useful to perform the mapping from flux to counts, which should be performed uniquely in each pixel $F = S / E_p$, using the mean exposure within small sub-regions.  Within a given sub-region, we map flux to counts using $F = S / \text{mean}(E)$, where the mean is taken over all pixels in the sub-region.  The number of sub-regions is given by \lstinline{nexp}, and all sub-regions have approximately the same area.  As \lstinline{nexp} approaches the number of pixels, the approximation becomes exact; however, for many applications the approximation converges for a relatively small number of exposure regions.  We recommend verifying, in any application, that results are stable as \lstinline{nexp} is increased.        

After configuring the \lstinline{NPTF} instance, the log-likelihood may be extracted, as a function of the model parameters, in addition to the prior range.  The log-likelihood and prior range may then be used with any external package for performing Bayesian or frequentist inference.  This is particularly useful if the user would like to combine likelihood functions between different energy bins or otherwise add to the default likelihood function, for example, incorporating nuisance parameters beyond those associated with individual templates.   The package \texttt{MultiNest}, however, is already incorporated into the \lstinline{NPTF} class and may be run immediately after configuring the \lstinline{NPTF} instance.  This is done simply by executing the command
\begin{lstlisting}
nptf.perform_scan(run_tag=None,nlive=100)
\end{lstlisting}
where \lstinline{nlive} is an integer that specifies the number of live points used in the sampling of the posterior distribution.  \texttt{MultiNest} recommends an \lstinline{nlive} $\sim$500-1000, though the parameter defaults to $100$ if unspecified for quick test runs.  Additional \texttt{MultiNest} arguments may be passed as a dictionary through the optional \lstinline{pymultinest_options} keyword (see the \href{http://nptfit.readthedocs.io}{online documentation} for more details).  The optional keyword \lstinline{run_tag} is used to create a sub-folder for the \texttt{MultiNest} output with that name.

After a scan has been run (or if a scan has been run previously and saved), the results may be loaded through the command
\begin{lstlisting}
nptf.load_scan(run_tag=None)
\end{lstlisting}
The \texttt{MultiNest} chains, which give a discretized view of the posterior distribution, may then be accessed through, for example, \lstinline{nptf.samples}.  An instance of the \texttt{PyMultiNest} analyzer class may be accessed through \lstinline{nptf.a}.  A small analysis package, described later in this section, is also provided for performing a few common analyses.  
 
\subsection*{ \lstinline{class NPTFit.psf_correction.PSFCorrection} }

This is the class used to construct the arrays \lstinline{f_ary} and \lstinline{df_rho_div_f_ary} for the PSF correction.  An instance of \lstinline{PSFCorrection} is initialized through 
\begin{lstlisting}
pc_inst = PSFCorrection.PSFCorrection(psf_dir=None, num_f_bins=10, n_psf=50000, n_pts_per_psf=1000, f_trunc=0.01, nside=128, psf_sigma_deg=None, delay_compute=False)
\end{lstlisting}
with keywords
\begin{widetext}
\begin{center}
\begin{tabular}{ |c|c|c|c |}
\hline
Argument &  Default & Purpose & \lstinline!type! \\ \hline \hline
\lstinline!psf_dir! & \lstinline!None!  & Where PSF arrays are stored  & \lstinline!str!  \\ \hline 
\lstinline!num_f_bins! & \lstinline!10! & Number of linear-spaced points in\lstinline!f_ary!  & \lstinline!int!  \\ \hline 
\lstinline!n_psf! & \lstinline!50000! & Number of MC simulations for determining \lstinline!df_rho_div_f_ary!  & \lstinline!int!  \\ \hline 
\lstinline!n_pts_per_psf! & \lstinline!1000! & Number of points drawn for each MC simulation & \lstinline!int! \\ \hline 
\lstinline!f_trunc! & \lstinline!0.01! & Minimum $f$ value & \lstinline!float!  \\ \hline 
\lstinline!nside! & \lstinline!128! & \lstinline!HEALPix! parameter for size of map  & \lstinline!int!  \\ \hline
\lstinline!psf_sigma_deg! & \lstinline!None! & Standard deviation $\sigma$ of 2-D Gaussian PSF  & \lstinline!float!  \\ \hline
\lstinline!delay_compute! & \lstinline!False! & If \lstinline!True!, PSF not Gaussian and will be specified later  & \lstinline!bool!  \\ \hline
\end{tabular}
\end{center}
\end{widetext}
Note that the arrays  \lstinline{f_ary} and \lstinline{df_rho_div_f_ary} depend both on the PSF of the detector as well as the pixelation of the data; at present the \lstinline{PSFCorrection} class requires the pixelation to be in the \texttt{HEALPix} pixelation.

The keyword \lstinline{psf_dir} points to the directory where the \lstinline{f_ary} and \lstinline{df_rho_div_f_ary} will be stored; if unspecified, they will be stored to the current directory.  The \lstinline{f_ary} consists of \lstinline{num_f_bins} entries linear spaced between $0$ and $1$.  The PSF correction involves placing many \mbox{(\lstinline{n_psf})} PSFs at random positions on the \texttt{HEALPix} map, drawing \lstinline{n_pts_per_psf} points from each PSF, and then looking at the distribution of points among pixels.  The larger \lstinline{n_psf} and \lstinline{n_pts_per_psf}, the more accurate the computation of \lstinline{df_rho_div_f_ary} will be.  However, the computation time of the PSF arrays also increases as these parameters are increased.  

By default the \lstinline{PSFCorrection} class assumes that the PSF is a 2-D Gaussian distribution:
\es{2DGaussian}{
\text{PSF}(r) = {1 \over 2 \pi \sigma^2} \exp\left[ - {r^2 \over 2 \sigma^2} \right] \,.
}     
Here, $\text{PSF}(r)$ describes the spread of arriving counts with angular distance $r$ away from the arrival direction.  The parameter \lstinline{psf_sigma_deg} denotes $\sigma$ in degrees.  Upon initializing \lstinline{PSFCorrection} with \lstinline{psf_sigma_deg} specified, the class automatically computes the array \lstinline{df_rho_div_f_ary} and stores it in the \lstinline{psf_dir} with a unique name related to the keywords.  If such a file already exists in the \lstinline{psf_dir}, then the code will simply load this file instead of recomputing it.  After initialization, the relevant arrays may be accessed by \lstinline{pc_inst.f_ary} and  \lstinline{pc_inst.df_rho_div_f_ary}. 

The \lstinline{PSFCorrection} class can also handle arbitrary PSF functions.  In this case, the class should be initialized with \lstinline{delay_compute = True}.  Then, the user should manually set the function \lstinline{pc_inst.psf_r_func} to the desired function $\text{PSF}(r)$.  This function will be discretized with \lstinline{pc_inst.psf_samples} points out to \lstinline{pc_inst.sample_psf_max} degrees from $r=0$.  These two quantities also need to be manually specified.  The user also needs to set \lstinline{pc_inst.psf_tag} to a string that will be used for saving the PSF arrays.  After these four attributes have been set manually by the user, the PSF arrays are computed and stored by executing \mbox{\lstinline{pc_inst.make_or_load_psf_corr()}}.

\subsection*{ \lstinline{def NPTFit.create_mask.make_mask_total} }
 
This function is used to make masks that can then be used to reduce the data and templates to a smaller ROI when performing the scan.  While these masks can always be made by hand, this function provides a simple masking interface for maps in the \texttt{HEALPix} format.  The \lstinline{make_mask_total} function can mask pixels by latitude, longitude, and radius from any point on the sphere.  See the \href{http://nptfit.readthedocs.io}{online documentation} for more specific examples.
 
\subsection*{ \lstinline{class NPTFit.dnds_analysis.Analysis} }
  
The analysis class may be used to extract useful information from the results of an NPTF performed using \texttt{MultiNest}.  The class also has built-in plotting features for making many of the most common types of visualizations for the parameter posterior distribution.  An instance of the analysis class can be instantiated by 
\begin{lstlisting}
an = Analysis(nptf, mask=None, pixarea=0.)
\end{lstlisting}
where \lstinline{nptf} is itself an instance of the \lstinline{NPTF} class that already has the results of a scan loaded.  The keyword arguments \lstinline{mask} and \lstinline{pixarea} are optional. The user should specify a \lstinline{mask} if the desired ROI for the analysis is different that that used in the scan.  The user should specify a \lstinline{pixarea} if the data is not in the \texttt{HEALPix} format.  The code will still assume the pixels are equal area with area \lstinline{pixarea}, which should be specified in sr.   

After initialization, the intensities of Poissonian and non-Poissonian templates, respectively, may be extracted from the analysis class by the commands 
\begin{lstlisting}
an.return_intensity_arrays_poiss(comp) 
\end{lstlisting}
and
\begin{lstlisting}
an.return_intensity_arrays_non_poiss(
    comp)
\end{lstlisting}
Here, \lstinline{comp} refers to the template key used by the Poissonian or non-Poissonian model.  The arrays returned give the mean intensities of that model in the ROI in units of  $\text{counts}\, \text{cm}^{-2}\text{s}^{-1}$, assuming the exposure map was in units of cm$^2$s.  The arrays computed over the full set of entries in the discretized posterior distribution output by \texttt{MultiNest}.  Thus, these intensity arrays may be interpreted as the 1-D posteriors for the intensities.  For additional keywords that may be used to customize the computation of the intensity arrays, see the \href{http://nptfit.readthedocs.io}{online documentation}.

The source-count distributions may also be accessed from the analysis class.  Executing
\begin{lstlisting}
an.return_dndf_arrays(comp, flux)
\end{lstlisting}
will return the discretized 1-D posterior distribution for $\text{mean}_\text{ROI} dN_p(F)/dF$ at flux $F$ for the PS model with template key \lstinline{comp}.  Note that the mean is computed over pixels $p$ in the ROI.  

The 1-D posterior distributions for the individual model parameters may be accessed by 
\begin{lstlisting}
A_poiss_post = an.return_poiss_parameter_posteriors(
    comp)
\end{lstlisting}
for Poissonian models, and 
\begin{lstlisting}
A_non_poiss_post, n_non_poiss_post, Sb_non_poiss_post = an.return_non_poiss_parameter_posteriors(comp)
\end{lstlisting}
for non-Poissonian models.  Here \lstinline{A_poiss_post} is a 1-D array of the discretized posterior distribution for the Poissonian template normalization parameter.  Similarly, \lstinline{A_non_poiss_post} is the posterior array for the non-Poissonian normalization parameter.  The arrays \lstinline{n_non_poiss_post} and \lstinline{Sb_non_poiss_post} are 2-D, where -- for example -- \mbox{\lstinline{n_non_poiss_post = [n_1_array, n_2_array, ...]}} and \lstinline{n_1_array} is a 1-D array for the posterior for $n_1$.   

Another useful piece of information that may be extracted from the scan is the Bayesian evidence:
\begin{lstlisting}
l_be, l_be_err = an.get_log_evidence()
\end{lstlisting}
returns the log of the Bayesian evidence along with the uncertainty on this estimate based on the resolution of the MCMC.

For information on the plotting capabilities in the analysis class, see the \href{http://nptfit.readthedocs.io}{online documentation} or the example in the following section.

\section{ \texttt{NPTFit}: an example}
\label{NPTFit-example}

In this section we give an example for how to perform an NPTF using \texttt{NPTFit}.  Many more examples are available in the \href{http://nptfit.readthedocs.io}{online documentation}.  This particular example reproduces aspects of the main results of~\cite{Lee:2015fea}, which found evidence for a spherical population of unresolved gamma-ray PSs around the Galactic Center.  The example uses the processed, public {\it Fermi} data made available with the release of the \texttt{NPTFit} package.  The data set consists of 413 weeks of \emph{Fermi} Pass 8 data in the UltracleanVeto event class (top quartile of events as ranked by PSF) from 2 to 20 GeV.  The map is binned in \texttt{HEALPix} with $\mathtt{nside} = 128$.  The data, along with the exposure map and background templates, may be downloaded from 
\begin{center}
\url{http://hdl.handle.net/1721.1/105492}.
\end{center}

In the example we will perform an NPTF on the sub-region where we mask the Galactic plane at latitude $|b| < 2^\circ$ and mask pixels with angular distance greater than $30^\circ$ from the Galactic Center.  We also mask identified PSs in the 3FGL PS catalog~\cite{Acero:2015hja} at 95\% containment using the provided PS mask, which is added to the geometric mask.  We include smooth templates for diffuse gamma-ray emission in the Milky Way (using the {\it Fermi} \texttt{p6v11} diffuse model), isotropic emission (which can also absorb instrumental backgrounds), and emission following the {\it Fermi} bubbles, which are taken to be uniform in flux following the spatial template in~\cite{Su:2010qj}.  We also include a dark matter template, which traces the line of sight integral of the square of a canonical NFW density profile.

We additionally include point source (non-Poissonian) models for the DM template, as well as for a disk template which corresponds to a doubly exponential thin-disk source distribution with scale height 0.3 kpc and radius 5 kpc. The source-count distributions for these are parameterized by singly-broken power laws, each described by four parameters $\{ A, F_{b,1}, n_1, n_2 \}$.

\subsection{Setting up the scan}

We begin the example by loading in the relevant modules, described in the previous section, that we will need to setup, perform, and analyze the scan.
\begin{lstlisting}
import numpy as np
# module for performing scan
from NPTFit import nptfit
# module for creating the mask
from NPTFit import create_mask as cm 
# module for determining the PSF correction
from NPTFit import psf_correction as pc 
# module for analyzing the output
from NPTFit import dnds_analysis 
\end{lstlisting}
Next, we create an instance of the \lstinline{NPTF} class, which is used to configure and perform a scan.
\begin{lstlisting}
n = nptfit.NPTF(tag='GCE_Example')
\end{lstlisting}
We assume here that the supplementary {\it Fermi} data has been downloaded to a directory \lstinline{'fermi_data'}.  Then, we may load in the data and exposure maps by 
\begin{lstlisting}
fermi_data = np.load('fermi_data/fermidata_counts.npy').astype(int)
fermi_exposure = np.load('fermi_data/fermidata_exposure.npy')
n.load_data(fermi_data, fermi_exposure)
\end{lstlisting}
Importantly, note that the exposure map has units of cm$^2$s.  Next, we use the \lstinline{create_mask} class to generate our ROI mask, which consists of both the geometric mask and the PS mask loaded in from the \lstinline{'fermi_data'} directory:
\begin{lstlisting}
pscmask=np.array(np.load('fermi_data/fermidata_pscmask.npy'), dtype=bool)
mask = cm.make_mask_total(band_mask = True, band_mask_range = 2, mask_ring = True, inner = 0, outer = 30, custom_mask = pscmask)
n.load_mask(mask)
\end{lstlisting}
The templates may also be loaded in from this directory,
\begin{lstlisting}
dif = np.load('fermi_data/template_dif.npy')
iso = np.load('fermi_data/template_iso.npy')
bub = np.load('fermi_data/template_bub.npy')
gce = np.load('fermi_data/template_gce.npy')
dsk = np.load('fermi_data/template_dsk.npy')
\end{lstlisting}
These templates are counts map (i.e. flux maps times the exposure map) that have been pre-smoothed by the PSF (except for the disk-correlated template labeled \lstinline{dsk}).  We then add them to our \lstinline{NPTF} instance with appropriately chosen keywords:
\begin{lstlisting}
n.add_template(dif, 'dif')
n.add_template(iso, 'iso')
n.add_template(bub, 'bub')
n.add_template(gce, 'gce')
n.add_template(dsk, 'dsk')

# remove the exposure correction for PS templates
rescale = fermi_exposure/np.mean(fermi_exposure)
n.add_template(gce/rescale, 'gce_np', units='PS')
n.add_template(dsk/rescale, 'dsk_np', units='PS')

\end{lstlisting}

Note that templates \lstinline{'gce_np'} and \lstinline{'dsk_np'} intended to be used in non-Poissonian models should trace the underlying PS distribution, without exposure correction, and are added with the keyword \lstinline{units='PS'}.

\subsection{Adding models}
Now that we have loaded in all of the external data and templates, we can add models to our \lstinline{NPTF} instance.  First, we add in the Poissonian models,
\begin{lstlisting}
n.add_poiss_model('dif', '$A_\mathrm{dif}$', False, fixed=True, fixed_norm=14.67)
n.add_poiss_model('iso', '$A_\mathrm{iso}$', [0,2], False)
n.add_poiss_model('gce', '$A_\mathrm{gce}$', [0,2], False)
n.add_poiss_model('bub', '$A_\mathrm{bub}$', [0,2], False)
\end{lstlisting}
All Poissonian models are taken to have linear priors, with prior ranges for the normalizations between 0 and 2.  However, the normalization of the diffuse background has been fixed to the value $14.67$, which is approximately the correct normalization in these units for this template, in order to provide an example of this syntax.  Next, we add in the two non-Poissonian models:
\begin{lstlisting}
n.add_non_poiss_model('gce_np', ['$A_\mathrm{gce}^\mathrm{ps}$','$n_1^\mathrm{gce}$','$n_2^\mathrm{gce}$','$S_b^{(1), \mathrm{gce}}$'], [[-6,1],[2.05,30],[-2,1.95],[0.05,40]], [True,False,False,False])
n.add_non_poiss_model('dsk_np', ['$A_\mathrm{dsk}^\mathrm{ps}$','$n_1^\mathrm{dsk}$','$n_2^\mathrm{dsk}$','$S_b^{(1), \mathrm{dsk}}$'], [[-6,1],[2.05,30],[-2,1.95],[0.05,40]], [True,False,False,False])
\end{lstlisting}
We have added in the models for disk-correlated and NFW-correlated (line of sight integral of the the NFW distribution squared) unresolved PS templates.  Each of these models takes singly-broken power-law source-count distributions.  In each case, the normalization parameter is taken to have a log-flat prior while the indices and breaks are taken to have linear priors.  The units of the breaks are specified in terms of counts. 

\subsection{Configure scan with PSF correction}
In this energy range and with this data set, the PSF may be modeled by a 2-D Gaussian distribution with $\sigma = 0.1812^\circ$.  From this, we are able to construct the PSF-correction arrays:\footnote{For an example of how to construct these arrays with a more complicated, non-Gaussian PSF function, see the \href{http://nptfit.readthedocs.io}{online documentation}.} 
\begin{lstlisting}
pc_inst = pc.PSFCorrection(psf_sigma_deg=0.1812)
f_ary, df_rho_div_f_ary = pc_inst.f_ary, pc_inst.df_rho_div_f_ary
\end{lstlisting}
These arrays are then passed into the \lstinline{NPTF} instance when we configure the scan:
\begin{lstlisting}
n.configure_for_scan(f_ary, df_rho_div_f_ary, nexp=1)
\end{lstlisting}
Note that since our ROI is relatively small and the exposure map does not change significantly over the region, we have a single exposure region with \lstinline{nexp=1}.

\subsection{Performing the scan with \texttt{MultiNest}}
We perform the scan using \texttt{MultiNest} with \lstinline{nlive=100} as an example to demonstrate the basic features and conclusions of this analysis while being able to perform the scan in a reasonable amount of time on a single processor, although ideally \lstinline{nlive} should be set to a higher value for more reliable results:
\begin{lstlisting}
n.perform_scan(nlive=100)
\end{lstlisting}

\subsection{Analyzing the results}
Now, we are ready to analyze the results of the scan.  First we load in relevant modules:
\begin{lstlisting}
import corner
import matplotlib.pyplot as plt
\end{lstlisting}
and then we load in the results of the scan (configured as above),
\begin{lstlisting}
n.load_scan()
\end{lstlisting}
The chains, giving a discretized view of the posterior distribution, may be accessed simply through the attribute \lstinline{n.samples}.  However, we will analyze the results by using the analysis class provided with \texttt{NPTFit}.  We make an instance of this class simply by
\begin{lstlisting}
an = dnds_analysis.Analysis(n)
\end{lstlisting}

\subsubsection{Make triangle plots}

\begin{figure*}[htb]
\leavevmode
\begin{center}
\includegraphics[width=.98\textwidth]{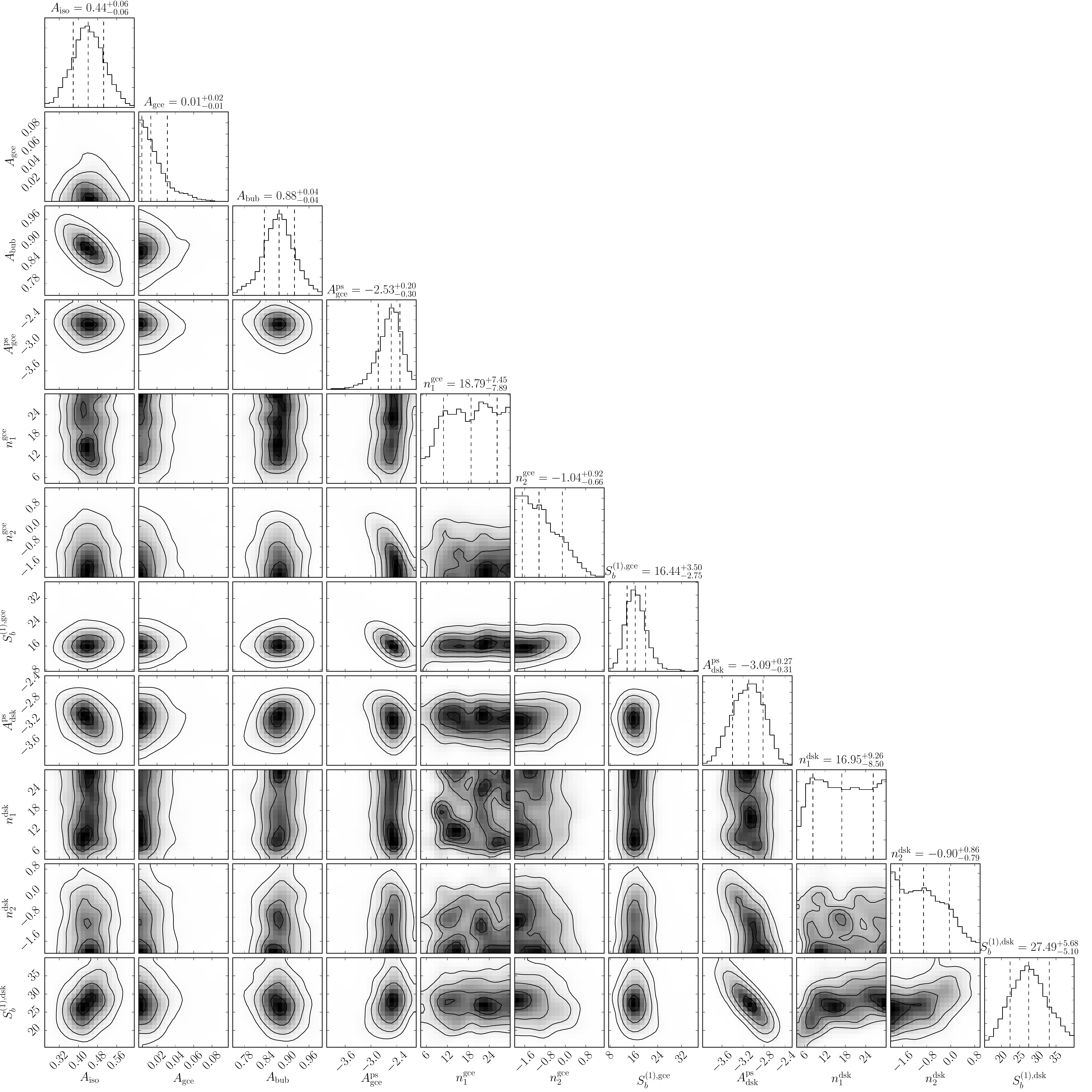}
\end{center}
\vspace{-.50cm}
\caption{The triangle plot obtained by analyzing the results of an NPTF in the Galactic Center, showing the one and two dimensional posteriors of the 11 parameters floated in the fit corresponding to three Poissonian and two non-Poissonian templates. For this analysis 3FGL point sources have been masked at 95\% containment. See text for details.}
\label{fig:gc_triangle}
\end{figure*}

Triangle plots are a simple and quick way of visualizing correlations in the posterior distribution.  Such plots may be generated through the command
\begin{lstlisting}
an.make_triangle()
\end{lstlisting}
which leads to the plot in Fig.~\ref{fig:gc_triangle}.

\subsubsection{Plot source-count distributions}

\begin{figure}[htb]
\leavevmode
\begin{center}
\includegraphics[width=.49\textwidth]{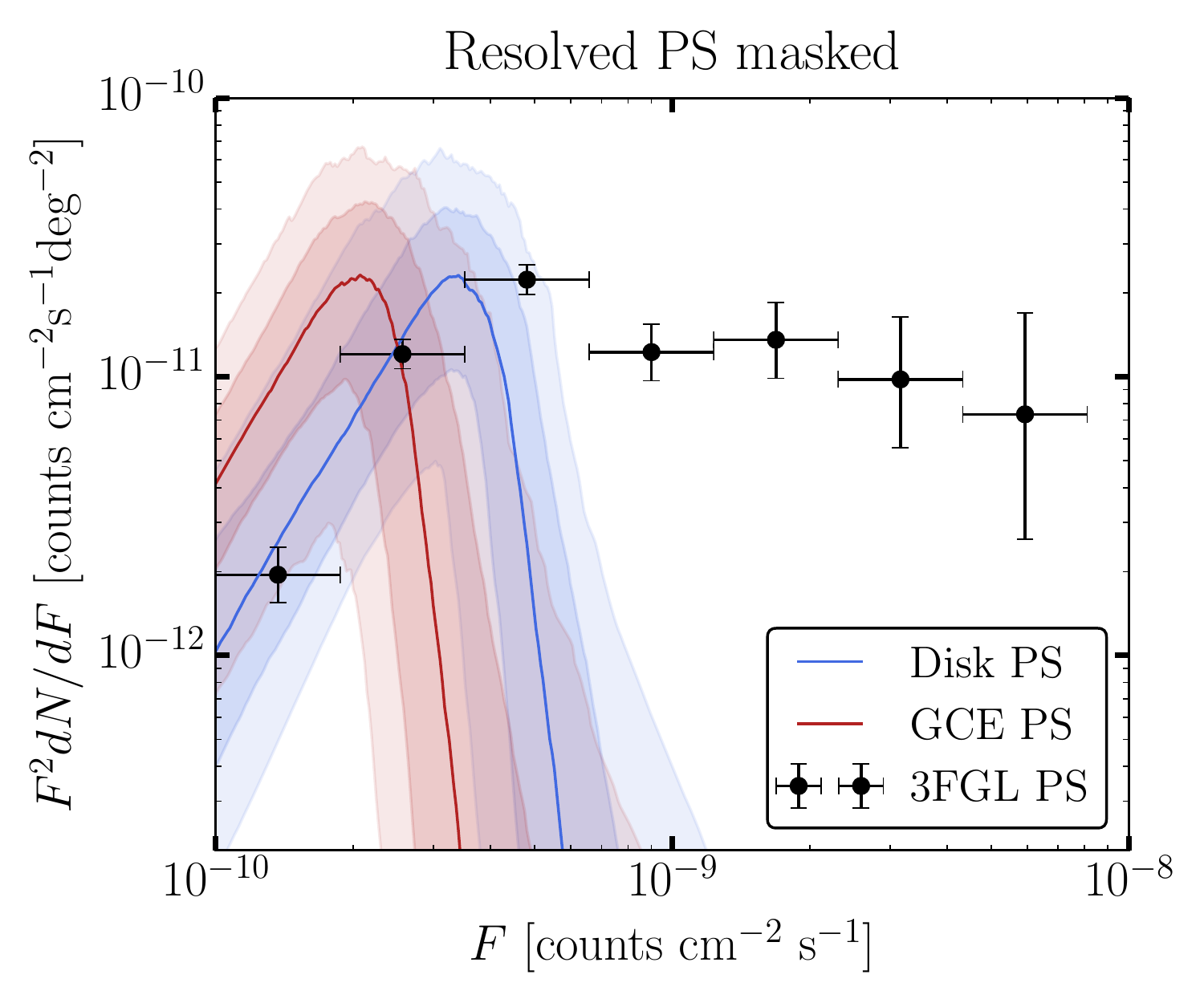}
\end{center}
\vspace{-.50cm}
\caption{The source-count distribution as constructed from the analysis class, for the example NPTF described in the main text.  This scan looks for  disk-correlated PSs along with PSs correlated with the expected DM template (GCE PSs).  Since all resolved PSs are masked in this analysis, the source-count distributions are seen to contribute dominantly below the 3FGL detection threshold.  A histogram of resolved 3FGL sources is also shown.}
\label{fig:gc_dndf}
\end{figure}

The source-count distributions for NFW- and disk-correlated point source models may be plotted with
\begin{lstlisting}
an.plot_source_count_median('dsk',smin=0.01,smax=1000,nsteps=1000,color='cornflowerblue',spow=2,label='Disk')
an.plot_source_count_band('dsk',smin=0.01,smax=1000,nsteps=1000,qs=[0.16,0.5,0.84],color='cornflowerblue',alpha=0.3,spow=2)
an.plot_source_count_median('gce',smin=0.01,smax=1000,nsteps=1000,color='forestgreen',spow=2,label='GCE')
an.plot_source_count_band('gce',smin=0.01,smax=1000,nsteps=1000,qs=[0.16,0.5,0.84],color='forestgreen',alpha=0.3,spow=2)
\end{lstlisting}
along with the following \texttt{matplotlib} plotting options. 
\begin{lstlisting}
plt.yscale('log')
plt.xscale('log')
plt.xlim([5e-11,5e-9])
plt.ylim([2e-13,1e-10])
plt.tick_params(axis='x', length=5, width=2, labelsize=18)
plt.tick_params(axis='y', length=5, width=2, labelsize=18)
plt.ylabel('$F^2 dN/dF$ [counts/cm$^2$/s/deg$^2$]', fontsize=18)
plt.xlabel('$F$  [counts/cm$^2$/s]', fontsize=18)
plt.title('Galactic Center NPTF', y=1.02)
plt.legend(fancybox=True)
plt.tight_layout()
\end{lstlisting}
This is shown in Fig.~\ref{fig:gc_dndf}. Contribution from both NFW- and disk-correlated PSs may be seen, with NFW-correlated sources contributing dominantly at lower flux values.  In that figure, we also show a histogram of the detected 3FGL sources within the relevant energy range and region, with vertical error bars indicating the 68\% confidence interval from Poisson counting uncertainties only.\footnote{The data for plotting these points is available in the \href{http://nptfit.readthedocs.io}{online documentation}.}  Since we have explicitly masked all 3FGL sources, we see that the disk- and NFW-correlated PS templates contribute at fluxes near and below the 3FGL PS detection threshold, which is $\sim$$5 \times 10^{-10}$ counts cm$^{-2}$ s$^{-1}$ in this case.

\subsubsection{Plot intensity fractions}

The intensity fractions for the smooth and PS NFW-correlated models may be plotted with

\begin{lstlisting}
an.plot_intensity_fraction_non_poiss('gce', bins=800, color='cornflowerblue', label='GCE PS')
an.plot_intensity_fraction_poiss('gce', bins=800, color='lightsalmon', label='GCE DM')
plt.xlabel('Flux fraction (%)')
plt.legend(fancybox = True)
plt.xlim(0,6)
\end{lstlisting}

\begin{figure}[htbp]
\centering
\includegraphics[width=0.45\textwidth]{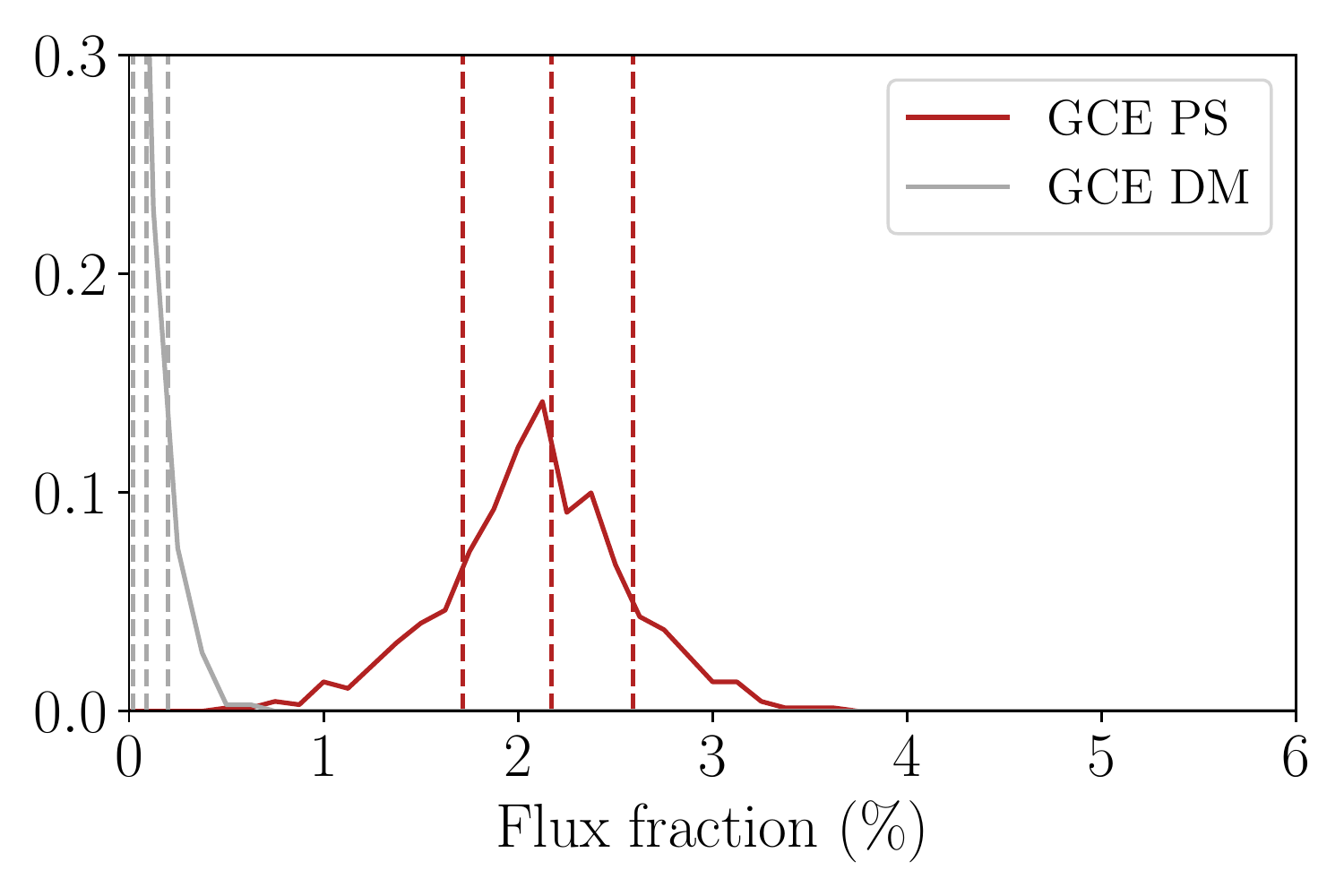} 
\caption{Intensity fractions for the smooth (green) and point source (red) templates correlating with the DM template, obtained by analyzing the results of an NPTF in the Galactic Center with 3FGL point sources masked at 95\% containment.}
\label{fig:gc_intensity}
\end{figure}

This is shown in Fig.~\ref{fig:gc_intensity}. We immediately see a preference for NFW-correlated point sources over the smooth NFW component.

\subsubsection{Further analyses}

The example above may easily be pushed further in many directions, many of which are outline in~\cite{Lee:2015fea}.  For example, a natural method for performing model comparison in the Bayesian framework is to compute the Bayes factor between two models.  Here, for example, we may compute the Bayes factor between the model with and without NFW-correlated PSs.  This involves repeating the scan described above but only adding in disk-correlated PSs.  Then, by comparing the global Bayesian evidence between the two scans (see Sec.~\ref{NPTFit-orientation} for the syntax on how to extract the Bayesian evidence), we find a Bayes factor $\sim$$10^3$ in preference for the model with spherical PSs.

\begin{figure}[htb]
\leavevmode
\begin{center}
\includegraphics[width=.49\textwidth]{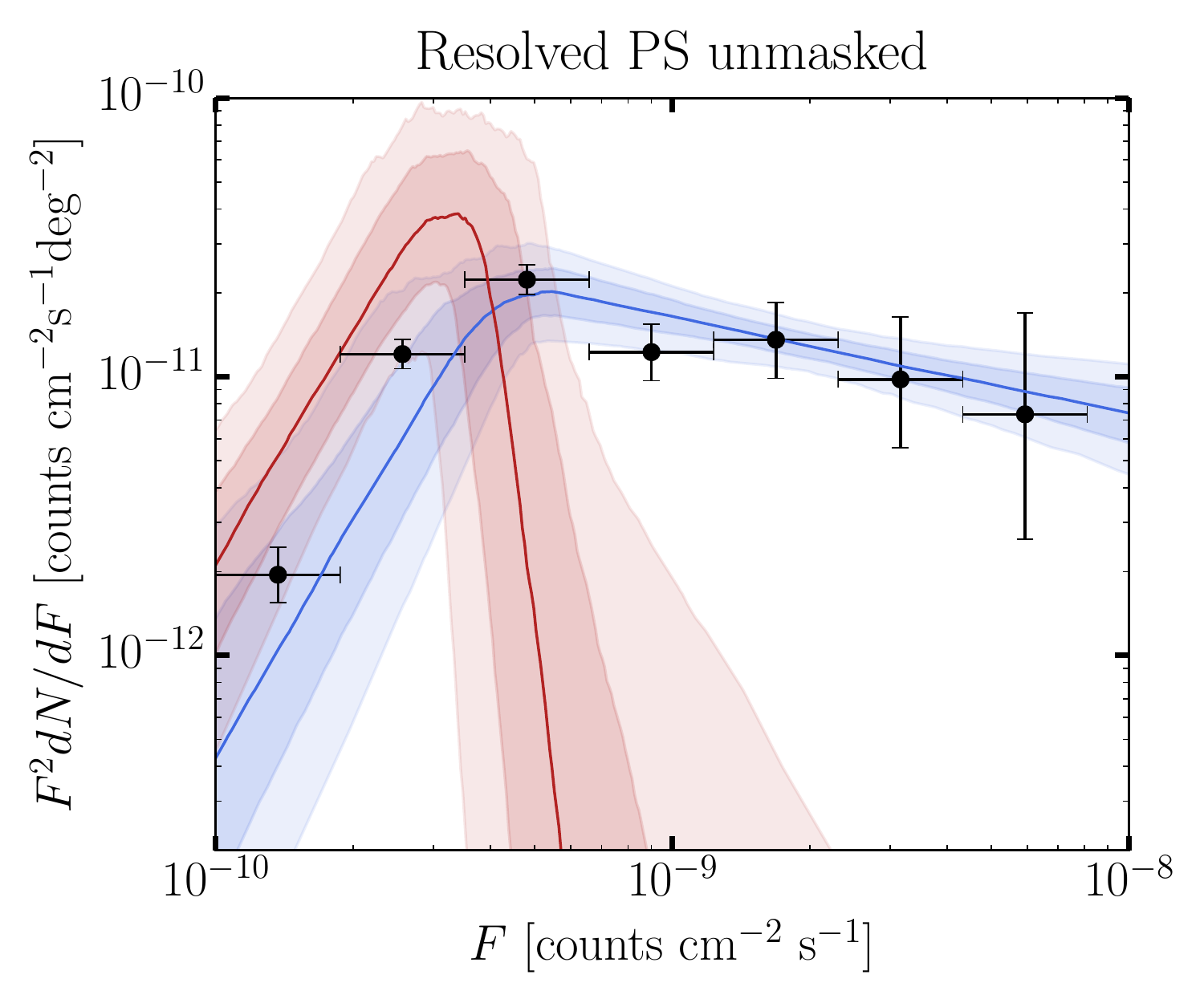}
\end{center}
\vspace{-.50cm}
\caption{As in Fig.~\ref{fig:gc_dndf}, but in this case the resolved 3FGL sources were not masked.  The disk-correlated template accounts for the majority of the resolved PS emission. }
\label{dNdF: unmasked}
\end{figure}   

Another straightforward generalization of the example described above is simply to leave out the PS mask, so that the NFW- and disk-correlated PS templates must account for both the resolved and unresolved PSs.  The likelihood evaluations take longer, in this case, since there are pixels with higher photon counts compared to the 3FGL-masked scan.  The result for the source-count distribution from this analysis is shown in Fig.~\ref{dNdF: unmasked}.  In this case, the disk-correlated PS template accounts for the resolved 3FGL sources, while the NFW-correlated PS template contributes at roughly the same flux range as in the 3FGL masked case.  The Bayes factor in preference for the model with NFW-correlated PSs over that without -- as described above -- is found to be $\sim$$10^{10}$ in this case.

\section{Conclusion}
\label{Conclusion}

We have presented an open-source code package for performing non-Poissonian template fits.   We strongly recommend referring to the \href{http://nptfit.readthedocs.io}{online documentation} -- which will be kept up-to-date -- in addition to this paper accompanying the initial release.  There are many way in which \texttt{NPTFit} can be improved in the future.  For one, the  \texttt{NPTFit} package only handles a single energy bin at a time.  In a later version of the code we plan to incorporate the ability to scan over multiple energy bins simultaneously.  Additionally, there are a few areas -- such as the evaluation of the incomplete gamma functions -- where the \texttt{cython} code may still be sped up.  Such improvements to the computational cost are relevant for analyses of large data sets with many model parameters.  Of course, we welcome additional suggestions for how we may improve the code and better adapt it to applications beyond the gamma-ray applications it has been used for so far. 

\section*{Acknowledgements}
\vspace{-10pt}
Foremost, we thank Samuel Lee for contributing significantly not just to the original conception of the NPTF but also to an early version of the \texttt{NPTFit} package.  We also thank Lina Necib for contributing to \texttt{NPTFit}.  Additionally, we thank our collaborators Tim Linden, Mariangela Lisanti, Tracy Slatyer, and Wei Xue, who worked with us on projects utilizing an early version of \texttt{NPTFit}. We thank Douglas Finkbeiner, Dan Hooper, Christoph Weniger, and Hannes Zechlin for discussions related to the NPTF and \texttt{NPTFit}.  NLR is supported in part by the American Australian Association's ConocoPhillips Fellowship. BRS is supported by a Pappalardo Fellowship in Physics at MIT.  The work of BRS was performed in part at the Aspen Center for Physics, which is supported by National Science Foundation grant PHY-1066293. This work is supported by the U.S. Department of Energy (DOE) under cooperative research agreement DE-SC-0012567 and DE-SC-0013999.

\appendix 

\section{Mathematical foundations of \texttt{NPTFit}}
\label{details}

In this section we present the mathematical foundation of the NPTF and the evaluation of the non-Poissonian likelihood in more detail that what was shown in Sec.~\ref{NPTF}. Note that many of the details presented in this section have appeared in the earlier works of \cite{Malyshev:2011zi,Lee:2014mza,Lee:2015fea}, however we have reproduced these here in order to have a single clear picture of the method.

The remainder of this section is divided as follows. Firstly we outline how to determine the generating functions for the Poissonian and non-Poissonian case. We then describe how we account for finite PSF corrections.

\subsection{The (non-)Poissonian generating function}

There are two reasons why the evaluation of the Poissonian likelihood for traditional template fitting can be evaluated rapidly. The first of these is that the functional form of the Poissonian likelihood is simple. Secondly, and more importantly, is the fact that if we have two discrete random variables $X$ and $Y$ that follow Poisson distributions with means $\mu_1$ and $\mu_2$, then the random variable $Z = X + Y$ again follows a Poisson distribution with mean $\mu_1 + \mu_2$. This generalizes to combining an arbitrary number of random Poisson distributed variables and is why we were able to write $\mu_{p,\ell}(\bm{\theta}) = A_{\ell}(\bm{\theta})T_{p,\ell}^{(S)}$ in Sec.~\ref{NPTF}. This fact is not true when combining arbitrary random variables, and in particular if we add in a template following non-Poissonian statistics. 

An elegant solution to this problem was introduced in~\cite{Malyshev:2011zi}, using the method of generating functions. As we are always dealing with pixelized maps containing discrete counts (of photons or otherwise), for any model of interest there will always be a discrete probability distribution $p_k$, the probability of observing $k=0, 1, 2, \ldots$ counts. In terms of these, we then define the probability generating function as in~\eqref{prob-gen}. The property of probability generating functions that make them so useful in the present context is as follows. Consider two random processes $X$ and $Y$, with generating functions $P_X(t)$ and $P_Y(t)$, that follow arbitrary and potentially different statistical distributions. Then the generating function of $Z = X + Y$ is simply given by the product $P_X(t) \cdot P_Y(t)$. In this subsection we will derive the appropriate form of $P(t)$ for Poissonian and non-Poissonian statistics.

To begin with, consider the purely Poissonian case. Here and throughout this section we consider only the likelihood in a single pixel; the likelihood over a full map is obtained from the product of the pixel-based likelihoods. Then for a Poisson distribution with an expected number of counts $\mu_p$ in a pixel $p$:
\begin{equation}
p_k = \frac{\mu_p^k e^{-\mu_p}}{k!}\,.
\end{equation}
Note that the variation of the $\mu_p$ across the full map will be a function of the model parameters, such that $\mu_p = \mu_p(\bm{\theta})$. In order to simplify the notation in this section however, we leave the $\bm{\theta}$ dependence implicit. Given the $p_k$ values, we then have:
\begin{equation}\begin{aligned}
P_{\rm P}(t) &= \sum_{k=0}^{\infty} \frac{\mu_p^k e^{-\mu_p}}{k!} t^k \\
&= e^{-\mu} \sum_{k=0}^{\infty} \frac{\left( \mu_p t \right)^k}{k!} \\
&= \exp \left[ \mu_p(t-1) \right]\,.
\label{eq:Pgen}
\end{aligned}\end{equation}
From this form, it is clear that if we have two Poisson distributions with means $\mu_p^{(1)}$ and $\mu_p^{(2)}$, the product of their generating functions will again describe a Poisson distribution, but with mean $\mu_p^{(1)} + \mu_p^{(2)}$.

Next we work towards the generating function in the non-Poissonian case. At the outset, we let $x_{p,m}$ denote the average number of sources in a pixel $p$ that emit exactly $m$ counts. In terms of this, the probability of finding $n_m$ $m$-count sources in this pixel is just a draw from a Poisson distribution with mean $x_{p,m}$, i.e.
\begin{equation}
p_{n_m} = \frac{x_{p,m}^{n_m} e^{-x_{p,m}}}{n_m!}\,.
\end{equation}
Given this, the probability to find $k$ counts from a population of $m$-count sources is
\begin{equation}
p_k^{(m)} = \left\{ \begin{array}{lc} p_{n_m}, & {\rm if}~k=m \cdot n_m~{\rm for~some~}n_m, \\ 0, & {\rm otherwise} \end{array} \right.\,.
\end{equation}
We can then use this to derive the non-Poissonian $m$-count generating function as follows:
\begin{equation}\begin{aligned}
P_{\rm NP}^{(m)}(t) &= \sum_k p_k t^k \\
&= \sum_{n_m} t^{m \cdot n_m} \frac{x_{p,m}^{n_m}e^{-x_{p,m}}}{n_m!} \\
&= \exp \left[ x_{p,m} (t^m - 1) \right]\,.
\end{aligned}\end{equation}
However this is just the generating function for $m$-count sources, to get the full non-Poissonian generating function we need to multiply this over all values of $m$. Doing so we arrive at
\begin{equation}\begin{aligned}
P_{\rm NP}(t) &= \prod_{m=1}^{\infty} \exp \left[ x_{p,m} (t^m - 1) \right] \\
&= \exp \left[ \sum_{m=1}^{\infty} x_{p,m} (t^m - 1) \right]\,,
\end{aligned}\end{equation}
justifying the form given in Sec.~\ref{NPTF}. Again recall for the full likelihood we can just multiply the pixel based likelihoods and that  $x_{p,m} = x_{p,m}(\bm{\theta})$.

So far we have said nothing of how to determine $x_{p,m}$, the average number of $m$-count source in pixel $p$. This value depends on the source-count distribution $dN_p/dS$, which specifies the distribution of sources as a function of their expected number of counts, $S$. Of course the physical object is $dN/dF$, where $F$ is the flux. This distinction was discussed in Sec.~\ref{NPTF}, and can be implemented in \texttt{NPTFit} to arbitrary precision. Nevertheless $dN_p/dS$ does not fully determine $x_{p,m}$ -- we need to account for the fact that a source that is expected to give $S$ photons could Poisson fluctuate to give $m$. As such any source can in principle contribute to $x_{p,m}$, and so integrating over the full distribution we arrive at:
\begin{equation}
x_{p,m} = \int_0^{\infty} dS \frac{dN_p}{dS}(S) \frac{S^m e^{-S}}{m!}\,.
\label{eq:xpm}
\end{equation}

An important part of implementing the NPTF in a rapid manner, which is a central feature of \texttt{NPTFit}, is the analytic evaluation of the integral in this equation. In order to do this, we need to have a specific form of the source-count distribution. For this purpose, we allow the source count distribution to be a multiply broken power-law and evaluate the integral for any number of breaks. The details of this calculation are presented in App.~\ref{xmcalc}. 

Putting the evaluation of the integral aside for the moment then, we have arrived at the full non-Poissonian generating function:
\begin{equation}\begin{aligned}
P_{\rm NP}(t) &= \exp \left[ \sum_{m=1}^{\infty} x_{p,m} (t^m - 1) \right]\,, \\
x_{p,m} &= \int_0^{\infty} dS \frac{dN_p}{dS}(S) \frac{S^m e^{-S}}{m!}\,.
\label{eq:NPgen}
\end{aligned}\end{equation}
Contrasting this with Eq.~\eqref{eq:Pgen}, we see that whilst the Poissonian likelihood is specified by a single number $\mu_p$, the non-Poissonian likelihood is instead specified by a distribution $dN_p/dS$.

In the case of multiple PS templates, we should multiply the independent probability generating functions.  However, this is equivalent to summing the $x_{p,m}$ parameters.  This is how multiple PS templates are incorporated into the \texttt{NPTFit} code:
\begin{equation}
x_{p,m} \to x_{p,m}^{\rm total} = \sum_{\ell=1}^{N_{\rm NPT}} x_{p,m}^{\ell}\,,
\end{equation}
where the sum over $\ell$ is over the contributions from individual PS templates. 

\subsection{Correcting for a finite point spread function}

The next factor to account for is the fact that in any realistic dataset there will be a non-zero PSF.  Here, we closely follow the discussion in~\cite{Malyshev:2011zi}. The PSF arises due to the inability of an instrument to perfectly reconstruct the original direction of the photon, neutrino, or quantity making up the counts. In practice, a finite PSF means that a source in one pixel can contribute counts to nearby pixels as well. To implement this correction, we modify the calculation of $x_{p,m}$ given in Eq.~\eqref{eq:NPgen}, which accounts for the distribution of sources as a function of $S$ and the fact that each one could Poisson fluctuate to give us $m$ counts. The finite PSF means that in addition to this, we also need to draw from the distribution $\rho(f)$, that determines the probability that a given source contributes a fraction of its flux $f$ in a given pixel. Once we know $\rho(f)$, this modifies our calculation of $x_{p,m}$ in Eq.~\eqref{eq:NPgen} -- now a source that is expected to contribute $S$ counts, will instead contribute $f S$, where $f$ is drawn from $\rho(f)$. As such we arrive at the result in~\eqref{xm-def}.

In \texttt{NPTFit} we determine $\rho(f)$ using Monte Carlo. To do this we place a number of PSs appropriately smeared by the PSF at random positions on a pixelized sphere. Then integrating over all pixels we can determine the fraction of the flux in each pixel $f_p$, $p=1,\ldots, N_{\rm pix}$, defined such that $f_1+f_2+\ldots = 1$. Note in practice one can truncate this sum at some minimal value of $f$ without impacting the argument below. From the set $\left\{f_p \right\}$, we then denote by $\Delta n(f)$ the number of fractions for $n$ point sources that fall within some range $\Delta f$. From these quantities, we may determine $\rho(f)$ as
\begin{equation}
\rho(f) = \lim_{\substack{\Delta f \to 0 \\ n \to \infty}} \frac{\Delta n(f)}{n \Delta f}\,,
\end{equation}
which is normalized such that $\int df~f \rho(f) = 1$. From this definition we see that the case of a vanishing PSF is just $\rho(f) = \delta(f-1)$ - i.e. the flux is always completely in the pixel with the PS.

\section{\texttt{NPTFit}: algorithms}
\label{algorithms}

The generating-function formalism for calculating the probabilities $p_{n_p}^{(p)}({\bm \theta})$ is described at the end of Sec.~\ref{NPTF} and in more detail in App.~\ref{details}.  In particular -- given the generating function $P(t)$ -- we are instructed to calculate the probabilities by taking $n_p$ derivatives as in~\eqref{deriv}.  However, taking derivatives is numerically costly, and so instead we have developed recursive algorithms for computing these probabilities.  In the same spirit, we analytically evaluate the $x_{p,m}$ parameters defined in~\eqref{xm-def} for the multiply-broken source-count distribution in order to facilitate a fast evaluation of the NPTF likelihood function.  In this section, we overview these methods that are essential to making \texttt{NPTFit} a practical software package.   

In general we may write the full single pixel generating function for a model containing an arbitrary number of Poissonian and non-Poissonian templates as:
\begin{equation}
P(t) = e^{f(t)}\,,
\end{equation}
where we have defined
\begin{equation}
f(t) \equiv \mu_p(t-1) + \sum_{m=1}^{\infty} x_{p,m} (t^m - 1)\,.
\end{equation}
Above, $x_{p,m} $ represents the average number of $m$-count source in pixel $p$. The remaining task is to efficiently calculate the probabilities $p_k$, which are formally defined in terms of derivatives through~\eqref{deriv}. Nevertheless, derivatives are slow to implement numerically, so we instead use a recursion relation to determine $p_k$ in terms of $p_{< k}$.

To begin with, note that
\begin{equation}
f^{(k)} \equiv \left. \frac{d^k}{dt^k} f(t) \right|_{t=0} = \left\{ \begin{array}{lc} -(\mu_p + \sum_{m=1}^{\infty} x_{p,m}), & k=0\,, \\ \mu_p + x_{p,1}, & k=1\,, \\ k! x_{p,k}, & k > 1\,. \end{array} \right.
\label{fk}
\end{equation}
For the rest of this discussion, we suppress the pixel index $p$, though one should keep in mind that this process must be performed independently in every pixel.
From~\eqref{fk}, we can immediately write down
\begin{equation}\begin{aligned}
p_0 &= e^{f^{(0)}}\,, \\
p_1 &= f^{(1)} e^{f^{(0)}}\,.
\end{aligned}\end{equation}
Given $p_0$ and $p_1$, we may write our recursion relation for $k > 1$ as
\es{recursion}{
p_k = \sum_{n=0}^{k-1} \frac{1}{k(k-n-1)!} f^{(k-n)} p_n\,,
}
which as mentioned requires the knowledge of all $p_{< k}$. 

To derive~\eqref{recursion}, we first define
\begin{equation}
F^{(k)}(t) \equiv \frac{d^k}{dt^k} e^{f(t)}\,.
\end{equation}
Then, for example,
\begin{equation}
F^{(1)}(t) = f^{(1)}(t) e^{f^{(0)}(t)}\,.
\end{equation}
From here to determine $F^{(k)}(t)$ we simply need $k-1$ more derivatives. Using the generalized Leibniz rule, we have
\begin{equation}\begin{aligned}
F^{(k)}(t) &= \frac{d^{k-1}}{dt^{k-1}} \left( f^{(1)}(t) e^{f^{(0)}(t)} \right) \\
&= \sum_{n=0}^{k-1} \begin{pmatrix} k-1 \\ n \end{pmatrix} \frac{d^{k-1-n}}{dt^{k-1-n}} f^{(1)}(t) \frac{d^n}{dt^n} e^{f^{(0)}(t)} \\
&= \sum_{n=0}^{k-1} \begin{pmatrix} k-1 \\ n \end{pmatrix} f^{(k-n)}(t) F^{(n)}(t)\,.
\end{aligned}\end{equation}
Then setting $t=0$ and recalling the definition of $p_k$, this yields
\begin{equation}\begin{aligned}
p_k &= \sum_{n=0}^{k-1} \frac{n!}{k!} \begin{pmatrix} k-1 \\ n \end{pmatrix} f^{(k-n)} p_n \\
&= \sum_{n=0}^{k-1} \frac{1}{k(k-n-1)!} f^{(k-n)} p_n\,,
\end{aligned}\end{equation}
as claimed.

To calculate the $f^{(k)}$ in a pixel $p$, we need to calculate the $x_{p,k}$ and the sum $\sum_{m=1}^\infty x_{p,m}$.  We may calculate these expressions analytically using the general source-count distribution in~\eqref{mbpl}.  To calculate the sums, we make use of the relation 
\begin{equation}\begin{aligned}
\sum_{m=1}^{\infty} x_{p,m} = &\int_0^{\infty} dS \frac{dN_p}{dS} e^{-S} \sum_{m=1}^{\infty} \frac{S^m}{m!} \\
= &\int_0^{\infty} dS \frac{dN_p}{dS} - \int_0^{\infty} dS \frac{dN_p}{dS} e^{-S} \\
= &\int_0^{\infty} dS \frac{dN_p}{dS} - x_{p,0}\,.
\end{aligned}
\label{eq:xmsum}
\end{equation}
Finiteness of the total flux, and also the probabilities, requires $n_1 > 2$ and $n_{k+1} < 2$.  However,  
both the integral and $x_{p,0}$, appearing in the last line above, may be divergent individually if $1 < n_{k+1} < 2$.  In this case, we analytically continue in $n_{k+1}$, evaluate the contributions individually, and then sum the two expressions to get a result that is finite across the whole range of allowable parameter space.  The expressions for the $x_{p,m}$ and the sums over these quantities are given in App.~\ref{xmcalc} in terms of incomplete gamma-functions.

\section{Analytic expressions for $x_{p,m}$ and $\sum_{m=1}^{\infty} x_{p,m}$}
\label{xmcalc}

In this appendix we derive analytic expressions for $x_{p,m}$ and $\sum_{m=1}^{\infty} x_{p,m}$, which  go into~\eqref{fk} and are needed to evaluate the non-Poissonian likelihood.  This is done by a straightforward application of~\eqref{eq:xpm} and~\eqref{eq:xmsum}. Recall that $x_{p,m} $ represents the average number of $m$-count source in pixel $p$. We begin by working explicitly through the 1- and 2-break source-count distributions before discussing the general case.

\subsection{1 break}
For a single break, the pixel-dependent source count distribution is given in terms of counts by
\begin{equation}
\frac{dN_p}{dS} = A \frac{T_p^{({\rm PS})}}{E_p} \left\{ \begin{array}{l} \left(S/S_b\right)^{-n_1},\;\;\;\; S \geq S_b \\ \left(S/S_b\right)^{-n_2},\;\;\;\; S < S_b \end{array} \right.\,.
\end{equation}
In the following, we will suppress the overall factor of $T_p^{({\rm PS})}/E_p$, since it does not play an important role in this discussion and may always be restored by simply rescaling $A$.  In the same spirit, we also suppress the pixel index $p$ in $x_{p,m}$.   

With this in mind, we may explicitly evaluate the expression for $x_{p,m}$ using~\eqref{eq:xpm}:
\begin{equation}\begin{aligned}
x_{m} =& \frac{A}{m!} \left[ S_b^{n_1} \int_{S_b}^{\infty} dS~S^{m-n_1} e^{-S} \right. \left. + S_b^{n_2} \int_0^{S_b} dS~S^{m-n_2} e^{-S} \right] \\
=& \frac{A}{m!} \left[ S_b^{n_1} \Gamma(1-n_1+m,S_b)+ S_b^{n_2} \Gamma(1-n_2+m) \right. \\
&\left.~~~~- S_b^{n_2} \Gamma(1-n_2+m,S_b) \right]\,.
\end{aligned}\end{equation}
Now using our general result~\eqref{eq:xmsum} above, we have
\begin{equation}\begin{aligned}
\sum_{m=1}^{\infty} x_{m} =& A \left[ S_b^{n_1} \int_{S_b}^{\infty} dS~S^{-n_1} + S_b^{n_2} \int_0^{S_b} dS~S^{-n_2} \right] - x_{p,0} \\
=&  A S_b \left[ \frac{1}{n_1-1} + \frac{1}{1-n_2} \right] - x_{0}\,.
\end{aligned}\end{equation}
This is useful because we already know $x_{0}$ from the general form of $x_{m}$ above.

\subsection{2 breaks}
For 2 breaks, the source-count distribution is given in terms of counts by
\begin{equation}
\frac{dN_p}{dS} = A \frac{T_p^{({\rm PS})}}{E_p} \left\{ \begin{array}{lc} \left( \frac{S}{S_{b,1}} \right)^{-n_1}, & S \geq S_{b,1} \\ \left(\frac{S}{S_{b,1}}\right)^{-n_2}, & S_{b,1} > S \geq S_{b,2} \\ \left( \frac{S_{b,2}}{S_{b,1}} \right)^{-n_2} \left(\frac{S}{S_{b,2}}\right)^{-n_3}, & S_{b,2} > S \end{array} \right.\,.
\end{equation}
Again suppressing the pixel-dependent pre-factors, an explicit evaluation gives
\begin{widetext}
\es{}{
\begin{aligned}
x_m &= \frac{A S_{b,1}^{n_1}}{ m!} \left[ \Gamma(1-n_1+m,S_{b,1}) + S_{b,1}^{n_2-n_1} \Gamma(1-n_2+m,S_{b,2}) - S_{b,1}^{n_2-n_1} \Gamma(1-n_2+m,S_{b,1}) \right. \\
&\left.~~~~~~~~~~~+ S_{b,1}^{n_2-n_1} S_{b,2}^{n_3-n_2} \Gamma(1-n_3+m) - S_{b,1}^{n_2-n_1} S_{b,2}^{n_3-n_2} \Gamma(1-n_3+m,S_{b,2}) \right] \,,
\end{aligned}
}
\begin{equation}\begin{aligned}
\sum_{m=1}^{\infty} x_m = AS_{b,1} &\left[ \frac{1}{n_1-1} + \frac{1}{1-n_2} \left( 1 - \left( \frac{S_{b,2}}{S_{b,1}} \right)^{1-n_2} \right)+\frac{1}{1-n_3} \left( \frac{S_{b,2}}{S_{b,1}} \right)^{1-n_2} \right] - x_0 \,.
\end{aligned}\end{equation}
\end{widetext}

\subsection{$k$ breaks}

The source-count distribution in the general $k$-break case is given in~\eqref{mbpl} in terms of flux.  In terms of counts and again suppressing pixel-dependent prefactors the result for $x_m$ and $\sum_{m=1}^{\infty} x_m$ is a simple generalization from the expressions for the $1$- and $2$-break cases:
\begin{widetext}
\es{}{
x_m = \frac{AS_{b,1}^{n_1}}{ m!} & \left[ \Gamma(1-n_1+m,S_{b,1}) 
+ \sum_{i=1}^{k-1} \left[ \prod_{j=1}^{i} S_{b,j}^{n_{j+1}-n_j} \right] \left\{\Gamma(1-n_{i+1}+m,S_{b,i+1}) - \Gamma(1-n_{i+1}+m,S_{b,i}) \right\}\right. \\
&\left.~+ \left[ \prod_{j=1}^{k} S_{b,j}^{n_{j+1}-n_j} \right] \left\{ \Gamma(1-n_{k+1}+m) - \Gamma(1-n_{k+1}+m,S_{b,k}) \right\}\right] \,,
}
\es{}{
\sum_{m=1}^{\infty} x_m = A S_{b,1} &\left[ \frac{1}{n_1-1} + \frac{1}{1-n_2} \left( 1 - \left( \frac{S_{b,2}}{S_{b,1}} \right)^{1-n_2} \right) \right. + \sum_{i=3}^{k} \frac{1}{1-n_i} \left[ \prod_{j=1}^{i-2} \left( \frac{S_{b,j+1}}{S_{b,j}} \right)^{1-n_{j+1}} \right] \left( 1 - \left( \frac{S_{b,i}}{S_{b,i-1}} \right)^{1-n_i} \right) \\
&\left.~+ \frac{1}{1-n_{k+1}} \left[ \prod_{j=1}^{k-1} \left( \frac{S_{b,j+1}}{S_{b,j}} \right)^{1-n_{j+1}} \right] \right] - x_0 \,.
}
\end{widetext}

\clearpage
\newpage
\cleardoublepage

\twocolumngrid
\def\bibsection{} 
\bibliographystyle{apsrev}
\clearpage
\cleardoublepage
\newpage
\bibliography{NPTF}

\begin{thebibliography}{60}
\expandafter\ifx\csname natexlab\endcsname\relax\def\natexlab#1{#1}\fi
\expandafter\ifx\csname bibnamefont\endcsname\relax
  \def\bibnamefont#1{#1}\fi
\expandafter\ifx\csname bibfnamefont\endcsname\relax
  \def\bibfnamefont#1{#1}\fi
\expandafter\ifx\csname citenamefont\endcsname\relax
  \def\citenamefont#1{#1}\fi
\expandafter\ifx\csname url\endcsname\relax
  \def\url#1{\texttt{#1}}\fi
\expandafter\ifx\csname urlprefix\endcsname\relax\def\urlprefix{URL }\fi
\providecommand{\bibinfo}[2]{#2}
\providecommand{\eprint}[2][]{\url{#2}}

\bibitem[{\citenamefont{Lee et~al.}(2015)\citenamefont{Lee, Lisanti, and
  Safdi}}]{Lee:2014mza}
\bibinfo{author}{\bibfnamefont{S.~K.} \bibnamefont{Lee}},
  \bibinfo{author}{\bibfnamefont{M.}~\bibnamefont{Lisanti}}, \bibnamefont{and}
  \bibinfo{author}{\bibfnamefont{B.~R.} \bibnamefont{Safdi}},
  \bibinfo{journal}{JCAP} \textbf{\bibinfo{volume}{1505}}, \bibinfo{pages}{056}
  (\bibinfo{year}{2015}), \eprint{1412.6099}.

\bibitem[{\citenamefont{Lee et~al.}(2016)\citenamefont{Lee, Lisanti, Safdi,
  Slatyer, and Xue}}]{Lee:2015fea}
\bibinfo{author}{\bibfnamefont{S.~K.} \bibnamefont{Lee}},
  \bibinfo{author}{\bibfnamefont{M.}~\bibnamefont{Lisanti}},
  \bibinfo{author}{\bibfnamefont{B.~R.} \bibnamefont{Safdi}},
  \bibinfo{author}{\bibfnamefont{T.~R.} \bibnamefont{Slatyer}},
  \bibnamefont{and} \bibinfo{author}{\bibfnamefont{W.}~\bibnamefont{Xue}},
  \bibinfo{journal}{Phys. Rev. Lett.} \textbf{\bibinfo{volume}{116}},
  \bibinfo{pages}{051103} (\bibinfo{year}{2016}), \eprint{1506.05124}.

\bibitem[{\citenamefont{Miyaji and Griffiths}(2002)}]{Miyaji:2001dp}
\bibinfo{author}{\bibfnamefont{T.}~\bibnamefont{Miyaji}} \bibnamefont{and}
  \bibinfo{author}{\bibfnamefont{R.~E.} \bibnamefont{Griffiths}},
  \bibinfo{journal}{Astrophys. J.} \textbf{\bibinfo{volume}{564}},
  \bibinfo{pages}{L5} (\bibinfo{year}{2002}), \eprint{astro-ph/0111393}.

\bibitem[{\citenamefont{Malyshev and Hogg}(2011)}]{Malyshev:2011zi}
\bibinfo{author}{\bibfnamefont{D.}~\bibnamefont{Malyshev}} \bibnamefont{and}
  \bibinfo{author}{\bibfnamefont{D.~W.} \bibnamefont{Hogg}},
  \bibinfo{journal}{Astrophys. J.} \textbf{\bibinfo{volume}{738}},
  \bibinfo{pages}{181} (\bibinfo{year}{2011}), \eprint{1104.0010}.

\bibitem[{\citenamefont{Goodenough and Hooper}(2009)}]{Goodenough:2009gk}
\bibinfo{author}{\bibfnamefont{L.}~\bibnamefont{Goodenough}} \bibnamefont{and}
  \bibinfo{author}{\bibfnamefont{D.}~\bibnamefont{Hooper}}
  (\bibinfo{year}{2009}), \eprint{0910.2998}.

\bibitem[{\citenamefont{Hooper and Goodenough}(2011)}]{Hooper:2010mq}
\bibinfo{author}{\bibfnamefont{D.}~\bibnamefont{Hooper}} \bibnamefont{and}
  \bibinfo{author}{\bibfnamefont{L.}~\bibnamefont{Goodenough}},
  \bibinfo{journal}{Phys.Lett.} \textbf{\bibinfo{volume}{B697}},
  \bibinfo{pages}{412} (\bibinfo{year}{2011}), \eprint{1010.2752}.

\bibitem[{\citenamefont{Boyarsky et~al.}(2011)\citenamefont{Boyarsky, Malyshev,
  and Ruchayskiy}}]{Boyarsky:2010dr}
\bibinfo{author}{\bibfnamefont{A.}~\bibnamefont{Boyarsky}},
  \bibinfo{author}{\bibfnamefont{D.}~\bibnamefont{Malyshev}}, \bibnamefont{and}
  \bibinfo{author}{\bibfnamefont{O.}~\bibnamefont{Ruchayskiy}},
  \bibinfo{journal}{Phys.Lett.} \textbf{\bibinfo{volume}{B705}},
  \bibinfo{pages}{165} (\bibinfo{year}{2011}), \eprint{1012.5839}.

\bibitem[{\citenamefont{Hooper and Linden}(2011)}]{Hooper:2011ti}
\bibinfo{author}{\bibfnamefont{D.}~\bibnamefont{Hooper}} \bibnamefont{and}
  \bibinfo{author}{\bibfnamefont{T.}~\bibnamefont{Linden}},
  \bibinfo{journal}{Phys.Rev.} \textbf{\bibinfo{volume}{D84}},
  \bibinfo{pages}{123005} (\bibinfo{year}{2011}), \eprint{1110.0006}.

\bibitem[{\citenamefont{Abazajian and Kaplinghat}(2012)}]{Abazajian:2012pn}
\bibinfo{author}{\bibfnamefont{K.~N.} \bibnamefont{Abazajian}}
  \bibnamefont{and}
  \bibinfo{author}{\bibfnamefont{M.}~\bibnamefont{Kaplinghat}},
  \bibinfo{journal}{Phys.Rev.} \textbf{\bibinfo{volume}{D86}},
  \bibinfo{pages}{083511} (\bibinfo{year}{2012}), \eprint{1207.6047}.

\bibitem[{\citenamefont{Hooper and Slatyer}(2013)}]{Hooper:2013rwa}
\bibinfo{author}{\bibfnamefont{D.}~\bibnamefont{Hooper}} \bibnamefont{and}
  \bibinfo{author}{\bibfnamefont{T.~R.} \bibnamefont{Slatyer}},
  \bibinfo{journal}{Phys.Dark Univ.} \textbf{\bibinfo{volume}{2}},
  \bibinfo{pages}{118} (\bibinfo{year}{2013}), \eprint{1302.6589}.

\bibitem[{\citenamefont{Gordon and Macias}(2013)}]{Gordon:2013vta}
\bibinfo{author}{\bibfnamefont{C.}~\bibnamefont{Gordon}} \bibnamefont{and}
  \bibinfo{author}{\bibfnamefont{O.}~\bibnamefont{Macias}},
  \bibinfo{journal}{Phys.Rev.} \textbf{\bibinfo{volume}{D88}},
  \bibinfo{pages}{083521} (\bibinfo{year}{2013}), \eprint{1306.5725}.

\bibitem[{\citenamefont{Abazajian et~al.}(2014)\citenamefont{Abazajian, Canac,
  Horiuchi, and Kaplinghat}}]{Abazajian:2014fta}
\bibinfo{author}{\bibfnamefont{K.~N.} \bibnamefont{Abazajian}},
  \bibinfo{author}{\bibfnamefont{N.}~\bibnamefont{Canac}},
  \bibinfo{author}{\bibfnamefont{S.}~\bibnamefont{Horiuchi}}, \bibnamefont{and}
  \bibinfo{author}{\bibfnamefont{M.}~\bibnamefont{Kaplinghat}},
  \bibinfo{journal}{Phys.Rev.} \textbf{\bibinfo{volume}{D90}},
  \bibinfo{pages}{023526} (\bibinfo{year}{2014}), \eprint{1402.4090}.

\bibitem[{\citenamefont{Daylan et~al.}(2016{\natexlab{a}})\citenamefont{Daylan,
  Finkbeiner, Hooper, Linden, Portillo, Rodd, and Slatyer}}]{Daylan:2014rsa}
\bibinfo{author}{\bibfnamefont{T.}~\bibnamefont{Daylan}},
  \bibinfo{author}{\bibfnamefont{D.~P.} \bibnamefont{Finkbeiner}},
  \bibinfo{author}{\bibfnamefont{D.}~\bibnamefont{Hooper}},
  \bibinfo{author}{\bibfnamefont{T.}~\bibnamefont{Linden}},
  \bibinfo{author}{\bibfnamefont{S.~K.~N.} \bibnamefont{Portillo}},
  \bibinfo{author}{\bibfnamefont{N.~L.} \bibnamefont{Rodd}}, \bibnamefont{and}
  \bibinfo{author}{\bibfnamefont{T.~R.} \bibnamefont{Slatyer}},
  \bibinfo{journal}{Phys. Dark Univ.} \textbf{\bibinfo{volume}{12}},
  \bibinfo{pages}{1} (\bibinfo{year}{2016}{\natexlab{a}}), \eprint{1402.6703}.

\bibitem[{\citenamefont{Calore et~al.}(2015)\citenamefont{Calore, Cholis, and
  Weniger}}]{Calore:2014xka}
\bibinfo{author}{\bibfnamefont{F.}~\bibnamefont{Calore}},
  \bibinfo{author}{\bibfnamefont{I.}~\bibnamefont{Cholis}}, \bibnamefont{and}
  \bibinfo{author}{\bibfnamefont{C.}~\bibnamefont{Weniger}},
  \bibinfo{journal}{JCAP} \textbf{\bibinfo{volume}{1503}}, \bibinfo{pages}{038}
  (\bibinfo{year}{2015}), \eprint{1409.0042}.

\bibitem[{\citenamefont{Abazajian et~al.}(2015)\citenamefont{Abazajian, Canac,
  Horiuchi, Kaplinghat, and Kwa}}]{Abazajian:2014hsa}
\bibinfo{author}{\bibfnamefont{K.~N.} \bibnamefont{Abazajian}},
  \bibinfo{author}{\bibfnamefont{N.}~\bibnamefont{Canac}},
  \bibinfo{author}{\bibfnamefont{S.}~\bibnamefont{Horiuchi}},
  \bibinfo{author}{\bibfnamefont{M.}~\bibnamefont{Kaplinghat}},
  \bibnamefont{and} \bibinfo{author}{\bibfnamefont{A.}~\bibnamefont{Kwa}},
  \bibinfo{journal}{JCAP} \textbf{\bibinfo{volume}{1507}}, \bibinfo{pages}{013}
  (\bibinfo{year}{2015}), \eprint{1410.6168}.

\bibitem[{\citenamefont{Ajello et~al.}(2016)}]{TheFermi-LAT:2015kwa}
\bibinfo{author}{\bibfnamefont{M.}~\bibnamefont{Ajello}} \bibnamefont{et~al.}
  (\bibinfo{collaboration}{Fermi-LAT}), \bibinfo{journal}{Astrophys. J.}
  \textbf{\bibinfo{volume}{819}}, \bibinfo{pages}{44} (\bibinfo{year}{2016}),
  \eprint{1511.02938}.

\bibitem[{\citenamefont{Macias et~al.}(2016)\citenamefont{Macias, Gordon,
  Crocker, Coleman, Paterson, Horiuchi, and Pohl}}]{Macias:2016nev}
\bibinfo{author}{\bibfnamefont{O.}~\bibnamefont{Macias}},
  \bibinfo{author}{\bibfnamefont{C.}~\bibnamefont{Gordon}},
  \bibinfo{author}{\bibfnamefont{R.~M.} \bibnamefont{Crocker}},
  \bibinfo{author}{\bibfnamefont{B.}~\bibnamefont{Coleman}},
  \bibinfo{author}{\bibfnamefont{D.}~\bibnamefont{Paterson}},
  \bibinfo{author}{\bibfnamefont{S.}~\bibnamefont{Horiuchi}}, \bibnamefont{and}
  \bibinfo{author}{\bibfnamefont{M.}~\bibnamefont{Pohl}}
  (\bibinfo{year}{2016}), \eprint{1611.06644}.

\bibitem[{\citenamefont{Clark et~al.}(2016)\citenamefont{Clark, Scott, Trotta,
  and Lewis}}]{Clark:2016mbb}
\bibinfo{author}{\bibfnamefont{H.~A.} \bibnamefont{Clark}},
  \bibinfo{author}{\bibfnamefont{P.}~\bibnamefont{Scott}},
  \bibinfo{author}{\bibfnamefont{R.}~\bibnamefont{Trotta}}, \bibnamefont{and}
  \bibinfo{author}{\bibfnamefont{G.~F.} \bibnamefont{Lewis}}
  (\bibinfo{year}{2016}), \eprint{1612.01539}.

\bibitem[{\citenamefont{Abazajian}(2011)}]{Abazajian:2010zy}
\bibinfo{author}{\bibfnamefont{K.~N.} \bibnamefont{Abazajian}},
  \bibinfo{journal}{JCAP} \textbf{\bibinfo{volume}{1103}}, \bibinfo{pages}{010}
  (\bibinfo{year}{2011}), \eprint{1011.4275}.

\bibitem[{\citenamefont{Hooper et~al.}(2013)\citenamefont{Hooper, Cholis,
  Linden, Siegal-Gaskins, and Slatyer}}]{Hooper:2013nhl}
\bibinfo{author}{\bibfnamefont{D.}~\bibnamefont{Hooper}},
  \bibinfo{author}{\bibfnamefont{I.}~\bibnamefont{Cholis}},
  \bibinfo{author}{\bibfnamefont{T.}~\bibnamefont{Linden}},
  \bibinfo{author}{\bibfnamefont{J.}~\bibnamefont{Siegal-Gaskins}},
  \bibnamefont{and} \bibinfo{author}{\bibfnamefont{T.~R.}
  \bibnamefont{Slatyer}}, \bibinfo{journal}{Phys.Rev.}
  \textbf{\bibinfo{volume}{D88}}, \bibinfo{pages}{083009}
  (\bibinfo{year}{2013}), \eprint{1305.0830}.

\bibitem[{\citenamefont{Calore et~al.}(2014)\citenamefont{Calore, Di~Mauro,
  Donato, and Donato}}]{Calore:2014oga}
\bibinfo{author}{\bibfnamefont{F.}~\bibnamefont{Calore}},
  \bibinfo{author}{\bibfnamefont{M.}~\bibnamefont{Di~Mauro}},
  \bibinfo{author}{\bibfnamefont{F.}~\bibnamefont{Donato}}, \bibnamefont{and}
  \bibinfo{author}{\bibfnamefont{F.}~\bibnamefont{Donato}},
  \bibinfo{journal}{Astrophys. J.} \textbf{\bibinfo{volume}{796}},
  \bibinfo{pages}{1} (\bibinfo{year}{2014}), \eprint{1406.2706}.

\bibitem[{\citenamefont{Cholis et~al.}(2015)\citenamefont{Cholis, Hooper, and
  Linden}}]{Cholis:2014lta}
\bibinfo{author}{\bibfnamefont{I.}~\bibnamefont{Cholis}},
  \bibinfo{author}{\bibfnamefont{D.}~\bibnamefont{Hooper}}, \bibnamefont{and}
  \bibinfo{author}{\bibfnamefont{T.}~\bibnamefont{Linden}},
  \bibinfo{journal}{JCAP} \textbf{\bibinfo{volume}{1506}}, \bibinfo{pages}{043}
  (\bibinfo{year}{2015}), \eprint{1407.5625}.

\bibitem[{\citenamefont{Petrovi{\'c} et~al.}(2015)\citenamefont{Petrovi{\'c},
  Serpico, and Zaharijas}}]{Petrovic:2014xra}
\bibinfo{author}{\bibfnamefont{J.}~\bibnamefont{Petrovi{\'c}}},
  \bibinfo{author}{\bibfnamefont{P.~D.} \bibnamefont{Serpico}},
  \bibnamefont{and}
  \bibinfo{author}{\bibfnamefont{G.}~\bibnamefont{Zaharijas}},
  \bibinfo{journal}{JCAP} \textbf{\bibinfo{volume}{1502}}, \bibinfo{pages}{023}
  (\bibinfo{year}{2015}), \eprint{1411.2980}.

\bibitem[{\citenamefont{Yuan and Ioka}(2015)}]{Yuan:2014yda}
\bibinfo{author}{\bibfnamefont{Q.}~\bibnamefont{Yuan}} \bibnamefont{and}
  \bibinfo{author}{\bibfnamefont{K.}~\bibnamefont{Ioka}},
  \bibinfo{journal}{Astrophys. J.} \textbf{\bibinfo{volume}{802}},
  \bibinfo{pages}{124} (\bibinfo{year}{2015}), \eprint{1411.4363}.

\bibitem[{\citenamefont{O'Leary et~al.}(2015)\citenamefont{O'Leary, Kistler,
  Kerr, and Dexter}}]{OLeary:2015gfa}
\bibinfo{author}{\bibfnamefont{R.~M.} \bibnamefont{O'Leary}},
  \bibinfo{author}{\bibfnamefont{M.~D.} \bibnamefont{Kistler}},
  \bibinfo{author}{\bibfnamefont{M.}~\bibnamefont{Kerr}}, \bibnamefont{and}
  \bibinfo{author}{\bibfnamefont{J.}~\bibnamefont{Dexter}},
  \bibinfo{journal}{1504.02477}  (\bibinfo{year}{2015}).

\bibitem[{\citenamefont{Brandt and Kocsis}(2015)}]{Brandt:2015ula}
\bibinfo{author}{\bibfnamefont{T.~D.} \bibnamefont{Brandt}} \bibnamefont{and}
  \bibinfo{author}{\bibfnamefont{B.}~\bibnamefont{Kocsis}},
  \bibinfo{journal}{Astrophys. J.} \textbf{\bibinfo{volume}{812}},
  \bibinfo{pages}{15} (\bibinfo{year}{2015}), \eprint{1507.05616}.

\bibitem[{\citenamefont{Linden et~al.}(2016)\citenamefont{Linden, Rodd, Safdi,
  and Slatyer}}]{Linden:2016rcf}
\bibinfo{author}{\bibfnamefont{T.}~\bibnamefont{Linden}},
  \bibinfo{author}{\bibfnamefont{N.~L.} \bibnamefont{Rodd}},
  \bibinfo{author}{\bibfnamefont{B.~R.} \bibnamefont{Safdi}}, \bibnamefont{and}
  \bibinfo{author}{\bibfnamefont{T.~R.} \bibnamefont{Slatyer}}
  (\bibinfo{year}{2016}), \eprint{1604.01026}.

\bibitem[{\citenamefont{Bartels et~al.}(2016)\citenamefont{Bartels,
  Krishnamurthy, and Weniger}}]{Bartels:2015aea}
\bibinfo{author}{\bibfnamefont{R.}~\bibnamefont{Bartels}},
  \bibinfo{author}{\bibfnamefont{S.}~\bibnamefont{Krishnamurthy}},
  \bibnamefont{and} \bibinfo{author}{\bibfnamefont{C.}~\bibnamefont{Weniger}},
  \bibinfo{journal}{Phys. Rev. Lett.} \textbf{\bibinfo{volume}{116}},
  \bibinfo{pages}{051102} (\bibinfo{year}{2016}), \eprint{1506.05104}.

\bibitem[{\citenamefont{Zechlin
  et~al.}(2016{\natexlab{a}})\citenamefont{Zechlin, Cuoco, Donato, Fornengo,
  and Vittino}}]{Zechlin:2015wdz}
\bibinfo{author}{\bibfnamefont{H.-S.} \bibnamefont{Zechlin}},
  \bibinfo{author}{\bibfnamefont{A.}~\bibnamefont{Cuoco}},
  \bibinfo{author}{\bibfnamefont{F.}~\bibnamefont{Donato}},
  \bibinfo{author}{\bibfnamefont{N.}~\bibnamefont{Fornengo}}, \bibnamefont{and}
  \bibinfo{author}{\bibfnamefont{A.}~\bibnamefont{Vittino}},
  \bibinfo{journal}{Astrophys. J. Suppl.} \textbf{\bibinfo{volume}{225}},
  \bibinfo{pages}{18} (\bibinfo{year}{2016}{\natexlab{a}}),
  \eprint{1512.07190}.

\bibitem[{\citenamefont{Ackermann et~al.}(2016)}]{TheFermi-LAT:2015ykq}
\bibinfo{author}{\bibfnamefont{M.}~\bibnamefont{Ackermann}}
  \bibnamefont{et~al.} (\bibinfo{collaboration}{Fermi-LAT}),
  \bibinfo{journal}{Phys. Rev. Lett.} \textbf{\bibinfo{volume}{116}},
  \bibinfo{pages}{151105} (\bibinfo{year}{2016}), \eprint{1511.00693}.

\bibitem[{\citenamefont{Zechlin
  et~al.}(2016{\natexlab{b}})\citenamefont{Zechlin, Cuoco, Donato, Fornengo,
  and Regis}}]{Zechlin:2016pme}
\bibinfo{author}{\bibfnamefont{H.-S.} \bibnamefont{Zechlin}},
  \bibinfo{author}{\bibfnamefont{A.}~\bibnamefont{Cuoco}},
  \bibinfo{author}{\bibfnamefont{F.}~\bibnamefont{Donato}},
  \bibinfo{author}{\bibfnamefont{N.}~\bibnamefont{Fornengo}}, \bibnamefont{and}
  \bibinfo{author}{\bibfnamefont{M.}~\bibnamefont{Regis}},
  \bibinfo{journal}{Astrophys. J.} \textbf{\bibinfo{volume}{826}},
  \bibinfo{pages}{L31} (\bibinfo{year}{2016}{\natexlab{b}}),
  \eprint{1605.04256}.

\bibitem[{\citenamefont{Lisanti et~al.}(2016)\citenamefont{Lisanti,
  Mishra-Sharma, Necib, and Safdi}}]{Lisanti:2016jub}
\bibinfo{author}{\bibfnamefont{M.}~\bibnamefont{Lisanti}},
  \bibinfo{author}{\bibfnamefont{S.}~\bibnamefont{Mishra-Sharma}},
  \bibinfo{author}{\bibfnamefont{L.}~\bibnamefont{Necib}}, \bibnamefont{and}
  \bibinfo{author}{\bibfnamefont{B.~R.} \bibnamefont{Safdi}}
  (\bibinfo{year}{2016}), \eprint{1606.04101}.

\bibitem[{\citenamefont{Daylan et~al.}(2016{\natexlab{b}})\citenamefont{Daylan,
  Portillo, and Finkbeiner}}]{Daylan:2016tia}
\bibinfo{author}{\bibfnamefont{T.}~\bibnamefont{Daylan}},
  \bibinfo{author}{\bibfnamefont{S.~K.~N.} \bibnamefont{Portillo}},
  \bibnamefont{and} \bibinfo{author}{\bibfnamefont{D.~P.}
  \bibnamefont{Finkbeiner}} (\bibinfo{year}{2016}{\natexlab{b}}),
  \eprint{1607.04637}.

\bibitem[{\citenamefont{Aartsen et~al.}(2013{\natexlab{a}})}]{Aartsen:2013bka}
\bibinfo{author}{\bibfnamefont{M.~G.} \bibnamefont{Aartsen}}
  \bibnamefont{et~al.} (\bibinfo{collaboration}{IceCube}),
  \bibinfo{journal}{Phys. Rev. Lett.} \textbf{\bibinfo{volume}{111}},
  \bibinfo{pages}{021103} (\bibinfo{year}{2013}{\natexlab{a}}),
  \eprint{1304.5356}.

\bibitem[{\citenamefont{Aartsen et~al.}(2013{\natexlab{b}})}]{Aartsen:2013jdh}
\bibinfo{author}{\bibfnamefont{M.~G.} \bibnamefont{Aartsen}}
  \bibnamefont{et~al.} (\bibinfo{collaboration}{IceCube}),
  \bibinfo{journal}{Science} \textbf{\bibinfo{volume}{342}},
  \bibinfo{pages}{1242856} (\bibinfo{year}{2013}{\natexlab{b}}),
  \eprint{1311.5238}.

\bibitem[{\citenamefont{Aartsen et~al.}(2015{\natexlab{a}})}]{Aartsen:2015knd}
\bibinfo{author}{\bibfnamefont{M.~G.} \bibnamefont{Aartsen}}
  \bibnamefont{et~al.} (\bibinfo{collaboration}{IceCube}),
  \bibinfo{journal}{Astrophys. J.} \textbf{\bibinfo{volume}{809}},
  \bibinfo{pages}{98} (\bibinfo{year}{2015}{\natexlab{a}}),
  \eprint{1507.03991}.

\bibitem[{\citenamefont{Aartsen et~al.}(2015{\natexlab{b}})}]{Aartsen:2015rwa}
\bibinfo{author}{\bibfnamefont{M.~G.} \bibnamefont{Aartsen}}
  \bibnamefont{et~al.} (\bibinfo{collaboration}{IceCube}),
  \bibinfo{journal}{Phys. Rev. Lett.} \textbf{\bibinfo{volume}{115}},
  \bibinfo{pages}{081102} (\bibinfo{year}{2015}{\natexlab{b}}),
  \eprint{1507.04005}.

\bibitem[{\citenamefont{Bechtol et~al.}(2015)\citenamefont{Bechtol, Ahlers,
  Di~Mauro, Ajello, and Vandenbroucke}}]{Bechtol:2015uqb}
\bibinfo{author}{\bibfnamefont{K.}~\bibnamefont{Bechtol}},
  \bibinfo{author}{\bibfnamefont{M.}~\bibnamefont{Ahlers}},
  \bibinfo{author}{\bibfnamefont{M.}~\bibnamefont{Di~Mauro}},
  \bibinfo{author}{\bibfnamefont{M.}~\bibnamefont{Ajello}}, \bibnamefont{and}
  \bibinfo{author}{\bibfnamefont{J.}~\bibnamefont{Vandenbroucke}}
  (\bibinfo{year}{2015}), \eprint{1511.00688}.

\bibitem[{\citenamefont{Murase and Waxman}(2016)}]{Murase:2016gly}
\bibinfo{author}{\bibfnamefont{K.}~\bibnamefont{Murase}} \bibnamefont{and}
  \bibinfo{author}{\bibfnamefont{E.}~\bibnamefont{Waxman}},
  \bibinfo{journal}{MNRAS}  (\bibinfo{year}{2016}), \eprint{1607.01601}.

\bibitem[{\citenamefont{Hasinger et~al.}(1993)\citenamefont{Hasinger, Burg,
  Giacconi, Hartner, Schmidt, Trumper, and Zamorani}}]{hasinger1993}
\bibinfo{author}{\bibfnamefont{G.}~\bibnamefont{Hasinger}},
  \bibinfo{author}{\bibfnamefont{R.}~\bibnamefont{Burg}},
  \bibinfo{author}{\bibfnamefont{R.}~\bibnamefont{Giacconi}},
  \bibinfo{author}{\bibfnamefont{G.}~\bibnamefont{Hartner}},
  \bibinfo{author}{\bibfnamefont{M.}~\bibnamefont{Schmidt}},
  \bibinfo{author}{\bibfnamefont{J.}~\bibnamefont{Trumper}}, \bibnamefont{and}
  \bibinfo{author}{\bibfnamefont{G.}~\bibnamefont{Zamorani}},
  \bibinfo{journal}{A\&A} \textbf{\bibinfo{volume}{275}}, \bibinfo{pages}{1}
  (\bibinfo{year}{1993}).

\bibitem[{\citenamefont{{Georgantopoulos}
  et~al.}(1993)\citenamefont{{Georgantopoulos}, {Stewart}, {Shanks},
  {Griffiths}, and {Boyle}}}]{1993MNRAS.262..619G}
\bibinfo{author}{\bibfnamefont{I.}~\bibnamefont{{Georgantopoulos}}},
  \bibinfo{author}{\bibfnamefont{G.~C.} \bibnamefont{{Stewart}}},
  \bibinfo{author}{\bibfnamefont{T.}~\bibnamefont{{Shanks}}},
  \bibinfo{author}{\bibfnamefont{R.~E.} \bibnamefont{{Griffiths}}},
  \bibnamefont{and} \bibinfo{author}{\bibfnamefont{B.~J.}
  \bibnamefont{{Boyle}}}, \bibinfo{journal}{MNRAS}
  \textbf{\bibinfo{volume}{262}}, \bibinfo{pages}{619} (\bibinfo{year}{1993}).

\bibitem[{\citenamefont{Gendreau et~al.}(1998)\citenamefont{Gendreau, Barcons,
  and Fabian}}]{Gendreau:1997di}
\bibinfo{author}{\bibfnamefont{K.~C.} \bibnamefont{Gendreau}},
  \bibinfo{author}{\bibfnamefont{X.}~\bibnamefont{Barcons}}, \bibnamefont{and}
  \bibinfo{author}{\bibfnamefont{A.~C.} \bibnamefont{Fabian}},
  \bibinfo{journal}{Mon. Not. Roy. Astron. Soc.}
  \textbf{\bibinfo{volume}{297}}, \bibinfo{pages}{41} (\bibinfo{year}{1998}),
  \eprint{astro-ph/9711083}.

\bibitem[{\citenamefont{Perri and Giommi}(2000)}]{Perri:2000fv}
\bibinfo{author}{\bibfnamefont{M.}~\bibnamefont{Perri}} \bibnamefont{and}
  \bibinfo{author}{\bibfnamefont{P.}~\bibnamefont{Giommi}},
  \bibinfo{journal}{Astron. Astrophys.} \textbf{\bibinfo{volume}{362}},
  \bibinfo{pages}{L57} (\bibinfo{year}{2000}), \eprint{astro-ph/0006298}.

\bibitem[{\citenamefont{Feyereisen et~al.}(2015)\citenamefont{Feyereisen, Ando,
  and Lee}}]{Feyereisen:2015cea}
\bibinfo{author}{\bibfnamefont{M.~R.} \bibnamefont{Feyereisen}},
  \bibinfo{author}{\bibfnamefont{S.}~\bibnamefont{Ando}}, \bibnamefont{and}
  \bibinfo{author}{\bibfnamefont{S.~K.} \bibnamefont{Lee}},
  \bibinfo{journal}{JCAP} \textbf{\bibinfo{volume}{1509}}, \bibinfo{pages}{027}
  (\bibinfo{year}{2015}), \eprint{1506.05118}.

\bibitem[{\citenamefont{Feyereisen et~al.}(2016)\citenamefont{Feyereisen,
  Tamborra, and Ando}}]{Feyereisen:2016fzb}
\bibinfo{author}{\bibfnamefont{M.~R.} \bibnamefont{Feyereisen}},
  \bibinfo{author}{\bibfnamefont{I.}~\bibnamefont{Tamborra}}, \bibnamefont{and}
  \bibinfo{author}{\bibfnamefont{S.}~\bibnamefont{Ando}}
  (\bibinfo{year}{2016}), \eprint{1610.01607}.

\bibitem[{\citenamefont{Behnel et~al.}(2011)\citenamefont{Behnel, Bradshaw,
  Citro, Dalcin, Seljebotn, and Smith}}]{behnel2010cython}
\bibinfo{author}{\bibfnamefont{S.}~\bibnamefont{Behnel}},
  \bibinfo{author}{\bibfnamefont{R.}~\bibnamefont{Bradshaw}},
  \bibinfo{author}{\bibfnamefont{C.}~\bibnamefont{Citro}},
  \bibinfo{author}{\bibfnamefont{L.}~\bibnamefont{Dalcin}},
  \bibinfo{author}{\bibfnamefont{D.}~\bibnamefont{Seljebotn}},
  \bibnamefont{and} \bibinfo{author}{\bibfnamefont{K.}~\bibnamefont{Smith}},
  \bibinfo{journal}{Computing in Science Engineering}
  \textbf{\bibinfo{volume}{13}}, \bibinfo{pages}{31 } (\bibinfo{year}{2011}),
  ISSN \bibinfo{issn}{1521-9615}.

\bibitem[{\citenamefont{Feroz et~al.}(2009)\citenamefont{Feroz, Hobson, and
  Bridges}}]{Feroz:2008xx}
\bibinfo{author}{\bibfnamefont{F.}~\bibnamefont{Feroz}},
  \bibinfo{author}{\bibfnamefont{M.~P.} \bibnamefont{Hobson}},
  \bibnamefont{and} \bibinfo{author}{\bibfnamefont{M.}~\bibnamefont{Bridges}},
  \bibinfo{journal}{Mon. Not. Roy. Astron. Soc.}
  \textbf{\bibinfo{volume}{398}}, \bibinfo{pages}{1601} (\bibinfo{year}{2009}),
  \eprint{0809.3437}.

\bibitem[{\citenamefont{Buchner et~al.}(2014)\citenamefont{Buchner,
  Georgakakis, Nandra, Hsu, Rangel, Brightman, Merloni, Salvato, Donley, and
  Kocevski}}]{Buchner:2014nha}
\bibinfo{author}{\bibfnamefont{J.}~\bibnamefont{Buchner}},
  \bibinfo{author}{\bibfnamefont{A.}~\bibnamefont{Georgakakis}},
  \bibinfo{author}{\bibfnamefont{K.}~\bibnamefont{Nandra}},
  \bibinfo{author}{\bibfnamefont{L.}~\bibnamefont{Hsu}},
  \bibinfo{author}{\bibfnamefont{C.}~\bibnamefont{Rangel}},
  \bibinfo{author}{\bibfnamefont{M.}~\bibnamefont{Brightman}},
  \bibinfo{author}{\bibfnamefont{A.}~\bibnamefont{Merloni}},
  \bibinfo{author}{\bibfnamefont{M.}~\bibnamefont{Salvato}},
  \bibinfo{author}{\bibfnamefont{J.}~\bibnamefont{Donley}}, \bibnamefont{and}
  \bibinfo{author}{\bibfnamefont{D.}~\bibnamefont{Kocevski}},
  \bibinfo{journal}{Astron. Astrophys.} \textbf{\bibinfo{volume}{564}},
  \bibinfo{pages}{A125} (\bibinfo{year}{2014}), \eprint{1402.0004}.

\bibitem[{\citenamefont{Feroz et~al.}(2013)\citenamefont{Feroz, Hobson,
  Cameron, and Pettitt}}]{Feroz:2013hea}
\bibinfo{author}{\bibfnamefont{F.}~\bibnamefont{Feroz}},
  \bibinfo{author}{\bibfnamefont{M.~P.} \bibnamefont{Hobson}},
  \bibinfo{author}{\bibfnamefont{E.}~\bibnamefont{Cameron}}, \bibnamefont{and}
  \bibinfo{author}{\bibfnamefont{A.~N.} \bibnamefont{Pettitt}}
  (\bibinfo{year}{2013}), \eprint{1306.2144}.

\bibitem[{\citenamefont{Feroz and Hobson}(2008)}]{Feroz:2007kg}
\bibinfo{author}{\bibfnamefont{F.}~\bibnamefont{Feroz}} \bibnamefont{and}
  \bibinfo{author}{\bibfnamefont{M.~P.} \bibnamefont{Hobson}},
  \bibinfo{journal}{Mon. Not. Roy. Astron. Soc.}
  \textbf{\bibinfo{volume}{384}}, \bibinfo{pages}{449} (\bibinfo{year}{2008}),
  \eprint{0704.3704}.

\bibitem[{\citenamefont{Skilling}(2006)}]{skilling2006}
\bibinfo{author}{\bibfnamefont{J.}~\bibnamefont{Skilling}},
  \bibinfo{journal}{Bayesian Anal.} \textbf{\bibinfo{volume}{1}},
  \bibinfo{pages}{833} (\bibinfo{year}{2006}),
  \urlprefix\url{http://dx.doi.org/10.1214/06-BA127}.

\bibitem[{\citenamefont{Gorski et~al.}(2005)\citenamefont{Gorski, Hivon,
  Banday, Wandelt, Hansen, Reinecke, and Bartelman}}]{Gorski:2004by}
\bibinfo{author}{\bibfnamefont{K.~M.} \bibnamefont{Gorski}},
  \bibinfo{author}{\bibfnamefont{E.}~\bibnamefont{Hivon}},
  \bibinfo{author}{\bibfnamefont{A.~J.} \bibnamefont{Banday}},
  \bibinfo{author}{\bibfnamefont{B.~D.} \bibnamefont{Wandelt}},
  \bibinfo{author}{\bibfnamefont{F.~K.} \bibnamefont{Hansen}},
  \bibinfo{author}{\bibfnamefont{M.}~\bibnamefont{Reinecke}}, \bibnamefont{and}
  \bibinfo{author}{\bibfnamefont{M.}~\bibnamefont{Bartelman}},
  \bibinfo{journal}{Astrophys. J.} \textbf{\bibinfo{volume}{622}},
  \bibinfo{pages}{759} (\bibinfo{year}{2005}), \eprint{astro-ph/0409513}.

\bibitem[{\citenamefont{P\'erez and Granger}(2007)}]{PER-GRA:2007}
\bibinfo{author}{\bibfnamefont{F.}~\bibnamefont{P\'erez}} \bibnamefont{and}
  \bibinfo{author}{\bibfnamefont{B.~E.} \bibnamefont{Granger}},
  \bibinfo{journal}{Computing in Science and Engineering}
  \textbf{\bibinfo{volume}{9}}, \bibinfo{pages}{21} (\bibinfo{year}{2007}),
  ISSN \bibinfo{issn}{1521-9615}, \urlprefix\url{http://ipython.org}.

\bibitem[{\citenamefont{Foreman-Mackey
  et~al.}(2016)\citenamefont{Foreman-Mackey, Vousden, Price-Whelan, Pitkin,
  Zabalza, Ryan, Emily, Smith, Ashton, Cruz
  et~al.}}]{dan_foreman_mackey_2016_53155}
\bibinfo{author}{\bibfnamefont{D.}~\bibnamefont{Foreman-Mackey}},
  \bibinfo{author}{\bibfnamefont{W.}~\bibnamefont{Vousden}},
  \bibinfo{author}{\bibfnamefont{A.}~\bibnamefont{Price-Whelan}},
  \bibinfo{author}{\bibfnamefont{M.}~\bibnamefont{Pitkin}},
  \bibinfo{author}{\bibfnamefont{V.}~\bibnamefont{Zabalza}},
  \bibinfo{author}{\bibfnamefont{G.}~\bibnamefont{Ryan}},
  \bibinfo{author}{\bibnamefont{Emily}},
  \bibinfo{author}{\bibfnamefont{M.}~\bibnamefont{Smith}},
  \bibinfo{author}{\bibfnamefont{G.}~\bibnamefont{Ashton}},
  \bibinfo{author}{\bibfnamefont{K.}~\bibnamefont{Cruz}}, \bibnamefont{et~al.},
  \emph{\bibinfo{title}{corner.py: corner.py v2.0.0}} (\bibinfo{year}{2016}),
  \urlprefix\url{https://doi.org/10.5281/zenodo.53155}.

\bibitem[{\citenamefont{Hunter}(2007)}]{Hunter:2007}
\bibinfo{author}{\bibfnamefont{J.~D.} \bibnamefont{Hunter}},
  \bibinfo{journal}{Computing In Science \& Engineering}
  \textbf{\bibinfo{volume}{9}}, \bibinfo{pages}{90} (\bibinfo{year}{2007}).

\bibitem[{\citenamefont{Johansson et~al.}(2013)}]{mpmath}
\bibinfo{author}{\bibfnamefont{F.}~\bibnamefont{Johansson}}
  \bibnamefont{et~al.}, \emph{\bibinfo{title}{mpmath: a {P}ython library for
  arbitrary-precision floating-point arithmetic (version 0.18)}}
  (\bibinfo{year}{2013}), \bibinfo{note}{{\tt http://mpmath.org/}}.

\bibitem[{\citenamefont{Galassi et~al.}(2015)\citenamefont{Galassi, Davies,
  Theiler, Gough, Jungman, Alken, Booth, Rossi, and Ulerich}}]{galassi2015gnu}
\bibinfo{author}{\bibfnamefont{M.}~\bibnamefont{Galassi}},
  \bibinfo{author}{\bibfnamefont{J.}~\bibnamefont{Davies}},
  \bibinfo{author}{\bibfnamefont{J.}~\bibnamefont{Theiler}},
  \bibinfo{author}{\bibfnamefont{B.}~\bibnamefont{Gough}},
  \bibinfo{author}{\bibfnamefont{G.}~\bibnamefont{Jungman}},
  \bibinfo{author}{\bibfnamefont{P.}~\bibnamefont{Alken}},
  \bibinfo{author}{\bibfnamefont{M.}~\bibnamefont{Booth}},
  \bibinfo{author}{\bibfnamefont{F.}~\bibnamefont{Rossi}}, \bibnamefont{and}
  \bibinfo{author}{\bibfnamefont{R.}~\bibnamefont{Ulerich}},
  \bibinfo{journal}{Library available online at http://www. gnu.
  org/software/gsl}  (\bibinfo{year}{2015}).

\bibitem[{\citenamefont{Oliphant}(2006)}]{oliphant2006guide}
\bibinfo{author}{\bibfnamefont{T.~E.} \bibnamefont{Oliphant}},
  \emph{\bibinfo{title}{A guide to NumPy}}, vol.~\bibinfo{volume}{1}
  (\bibinfo{publisher}{Trelgol Publishing USA}, \bibinfo{year}{2006}).

\bibitem[{\citenamefont{Acero et~al.}(2015)}]{Acero:2015hja}
\bibinfo{author}{\bibfnamefont{F.}~\bibnamefont{Acero}} \bibnamefont{et~al.}
  (\bibinfo{collaboration}{Fermi-LAT}) (\bibinfo{year}{2015}),
  \eprint{1501.02003}.

\bibitem[{\citenamefont{Su et~al.}(2010)\citenamefont{Su, Slatyer, and
  Finkbeiner}}]{Su:2010qj}
\bibinfo{author}{\bibfnamefont{M.}~\bibnamefont{Su}},
  \bibinfo{author}{\bibfnamefont{T.~R.} \bibnamefont{Slatyer}},
  \bibnamefont{and} \bibinfo{author}{\bibfnamefont{D.~P.}
  \bibnamefont{Finkbeiner}}, \bibinfo{journal}{Astrophys.J.}
  \textbf{\bibinfo{volume}{724}}, \bibinfo{pages}{1044} (\bibinfo{year}{2010}),
  \eprint{1005.5480}.

\end{thebibliography}

\end{document}